\documentclass[apj]{emulateapj}

\usepackage{natbib}
\usepackage{ifpdf}
\usepackage[colorlinks=true,linkcolor=blue,allcolors=blue]{hyperref}

\usepackage{graphicx}
\usepackage{amssymb}
\usepackage[caption=false]{subfig}
\newcommand\msun{{\,M_\odot}}

\newcommand\zsun{{\rm \,Z_\odot}}
\newcommand\lsun{{\rm \,L_\odot}}

\usepackage{aecompl}
\usepackage[T1]{fontenc}
\usepackage{threeparttable}
\usepackage{times}
\usepackage{amssymb}

\begin{document}
\shorttitle{DLAs in dwarf galaxies}
\shortauthors{JEON ET AL.}

\title{Probing the assembly of dwarf galaxies through cosmic time with damped Lyman-$\alpha$ absorption spectroscopy}
\author{Myoungwon Jeon\altaffilmark{1,2},
            Gurtina Besla\altaffilmark{2}
            Volker Bromm\altaffilmark{3}
            }

\altaffiltext{1}{Center for Global Converging Humanities, Kyung Hee University, Republic of Korea; myjeon@khu.ac.kr}
\altaffiltext{2}{Department of Astronomy, University of Arizona, 933 North Cherry Avenue, Tucson, AZ 85721, USA }
\altaffiltext{3}{Department of Astronomy, University of Texas, TX 78712, USA}

\begin{abstract}
We investigate the absorption features associated with a gas-rich dwarf galaxy using cosmological hydrodynamics simulations. Our goal is to explore whether the progenitors of the lowest mass dwarf galaxies known to harbor neutral hydrogen today ($M_{\ast}\approx10^6\msun$, $M_{\rm halo}=4\times10^9\msun$) could possibly be detected as Damped Lyman-$\alpha$ Absorbers (DLAs) over cosmic time. We trace the evolution of a single dwarf galaxy, pre-selected to contain DLAs, from the era of the first metal-free, so-called Population~III (Pop~III), stars, down to $z=0$, thus allowing us to study the metal enrichment history of DLAs associated with the simulated galaxy. We find that the progenitors of the simulated dwarf are expected to be seen for most of their evolution as DLAs that are contaminated by normal, Population~II, stars. The time period during which DLAs are only metal-enriched by Pop~III stars, on the other hand, is likely very brief, confined to high redshifts, $z\gtrsim6$. The susceptibility of the dwarfs to the external UV radiation background allows them to preserve neutral gas only at the centre (a few $\sim100$ pc). This results in a small probability that the simulated dwarf would be observed as a DLA. This study suggests that DLAs are unlikely to be hosted in the lowest mass dwarfs that can harbor neutral gas ($M_{\rm halo}\gtrsim 4\times10^9\msun$), below which neutral gas is unlikely to exist. However, this study does illustrate that, when detected, absorption lines provide a powerful method for probing ISM conditions inside the smallest dwarf galaxies at intermediate to high redshifts.

\end{abstract}


\section{Introduction}
The most common dwarf galaxies, with stellar masses $M_{\ast}\lesssim10^7\msun$ and virial masses $M_{\rm vir}\lesssim10^9\msun$, are of fundamental importance to our understanding of galaxy formation (see \citealp{Tolstoy2009} for a review). Their intrinsic faintness (with optical luminosities L$_V\lesssim10^6\lsun$) limits the direct observation of their stellar component to nearby sources. To gain insight into the properties and evolution of such galaxies, an alternative observational strategy is thus needed. In particular, probing the absorption features associated with dwarf galaxies enables us to investigate their gas conditions and on-going star formation at different cosmic times. We can thus re-construct their star formation histories (SFHs), to be compared with the predictions from increasingly realistic numerical simulations.

It has long been suggested that damped Lyman-$\alpha$ absorption (DLA) systems are vital tools to explore the gaseous conditions in and around the associated galaxies (for a review see \citealp{Wolfe2005, Krogager2017}). DLAs are generally defined as gas clouds with neutral hydrogen column density exceeding $N_{\rm H\,I}=10^{20.3} \rm cm^{-2}$ through the detection of damped Ly$\alpha$ absorption lines in quasar spectra (e.g., \citealp{Wolfe1986}). The nature of DLAs has been extensively discussed in previous studies particularly regarding the questions: which parts of a galaxy's environment, or which types of galaxies represent DLA systems? For instance, \citet{Prochaska1997} have initially suggested that DLAs originate from a massive disk galaxy, while \citet{Haehnelt1998} have proposed that the properties of the observed DLAs can be explained by infalling small systems during the galaxy assembly process. Some studies suggest that DLAs might be associated with a specific galaxy class such as actively star-forming galaxies at high-redshifts, classified as Lyman-$\alpha$ emitters (LAEs; e.g., \citealp{Fynbo2001, Fynbo2003, Rauch2008, Krogager2013, Noterdaeme2014}), or more massive galaxies with stellar mass of $\sim10^{10}\msun$, selected as Lyman-break galaxies (LBGs; e.g., \citealp{Moller2002, Verhamme2008, Steidel2010}).

A more direct way of probing the underlying origin of DLAs is to find their galaxy host counterparts via emission (e.g., \citealp{Moller2004, Fynbo2010, Fynbo2011, Fynbo2013, Krogager2013, Jorgenson2014, Zafar2017}). The flux-limited search, however, is likely to be biased to observe bright galaxies, thus missing faint, low-mass galaxies. Instead, absorption-based DLA observations are not hampered by the flux of galaxies, allowing one to explore a broad range of galaxy populations (e.g., \citealp{Fynbo1999, Fumagalli2015}). Assuming that the luminosity correlates with metallicity (\citealp{Ledoux2006, Fynbo2008, Christensen2014}), recently \citet{Krogager2017} have confirmed that DLAs appear to trace a broad range of galaxy mass, luminosities, and star formation properties. The authors have argued that the X-shooter campaign targets metal-rich DLAs. Its high detection rate of DLA counterparts, compared to blind surveys, thus suggests that the paucity of counterparts at low metallicity may be attributed to the inability of current facilities to detect faint galaxies.


To understand conditions in the pristine, early Universe, metal-poor DLAs are of particular interest, as they might either be associated with dwarf galaxies that have barely evolved chemically, or with infalling cold gas streams that feed galaxies at high-$z$ (e.g., \citealp{Pettini2008, Cooke2011, Cooke2015}). Vigorous theoretical efforts based on cosmological hydrodynamic simulations have been made in order to study the origin of DLAs (e.g., \citealp{Katz1996, Prochaska1997, Gardner1997, Gardner2001, Haehnelt1998, Nagamine2007, Pontzen2008, Tescari2009, Hong2010, Cen2012, Rahmati2014, Bird2015, Yuan2016, Berry2016}). The focus, however, has been on DLAs associated with rather massive galaxies (virial masses of $M_{\rm vir}\gtrsim10^{10-11}\msun$). In this study, we investigate the absorption features of gas associated with the assembly history of dwarf galaxies. We analyze a gas-rich dwarf galaxy, formed within a cosmological hydrodynamics, zoom-in simulation ($M_{\rm vir}=4\times10^9$, $M_{\ast}=8.8\times10^5\msun$, $M_{\rm gas}=4\times10^7\msun$ at $z=0$), where we trace cosmic evolution starting from the era of the first generation of stars, the so-called Population~III (Pop~III), down to $z=0$. We refer the reader to our previous work (\citealp{Jeon2017}) where the SFH of this dwarf has been analyzed in detail. 

The main goal of this work is to examine whether the progenitors of local, gas-rich dwarf galaxies, e.g. analogs of present day galaxies like Leo P and Leo A, could possibly be detected as DLA systems over cosmic time. In particular, we explore the questions: What do such possible absorption signatures look like, and what can the metal abundances inferred from DLAs tell us about the metal-enrichment history of a dwarf galaxy? Specifically, since we explicitly consider the contribution of heavy element enrichment from the first stars to the build-up of metals in dwarf galaxies, we can test the scenario that very metal-poor DLAs could contain the unique signature of Pop~III nucleosynthesis (e.g., \citealp{Frebel2007, Salvadori2007, Cooke2011, Salvadori2012, Kulkarni2013, Webster2015b, Cooke2017}).

One of the key challenges in elucidating early cosmic evolution is to understand the physical nature of the first generation of stars, formed a few hundred million years after the Big Bang (see, e.g., \citealp{Bromm2013} for a review). Due to the dramatic cosmological flux dimming, directly detecting individual Pop~III stars during their short lifetimes, $\lesssim 10$~Myr, is out of reach even for the upcoming next generation of ground-based extremely large telescopes, as well as for the {\it James Webb Space Telescope (JWST)}. 
Alternatively, one possible way of directly detecting Pop~III stars is through their death, if they die in very luminous explosions via a pair-instability supernova (PISN) with energies of $10^{51-53}$ erg (e.g., \citealp{Whalen2013, Hartwig2018}), or other types of superluminous supernovae (SLSNe) with similar extreme explosion energies (e.g., \citealp{Umeda2003}). Such difficulty in directly detecting Pop~III stars with existing and upcoming telescopes renders metal-poor DLAs possible alternative probes to indirectly search for the signatures of the first stars.

Recently, \citet{Cooke2017} have reported the discovery of extremely metal-poor DLAs with iron abundance of $\rm[Fe/H]<-2.81$ at $z=3.1$, using the high-resolution Keck HIRES spectrograph. Comparing the observed DLA abundance patterns with the nucleosynthetic yields of metal-free stars, they have suggested that the chemistry of the discovered DLA gas is consistent with the yields from a $20.5 \msun$ Pop~III star that died as a core-collapse supernova (CCSN). Interestingly, they inferred a large column density, $N_{\rm H\,I}=10^{20.32} \rm cm^{-2}$, which is 
sufficiently high to trigger star formation, leading them to propose that this system might be an immediate precursor to the formation of a first galaxy, or an ancestor of the local dwarf galaxies. Whether DLA systems continue to retain clear Pop~III signatures would depend on when the transition in star formation mode is achieved from Pop~III 
to the second, Population~II (Pop~II), generation of metal-poor stars.  Typically, this transition takes place at very early epochs, $z\gtrsim12$, depending on the mass of a host halo, such that more massive systems are likely to revert to Pop~II star formation at earlier times. Once the transition is accomplished, the Pop~III enrichment signature will be rapidly erased by the contribution from second generation stars \citep[e.g.][]{Ji2015}.

Another key factor governing the connection between gas in dwarfs and observed DLAs is whether the gravitational potential of the host halo can hold onto neutral gas, instead of allowing it to be blown away by stellar feedback and the external ultraviolet (UV) radiation background. Observations show that local ultra-faint dwarf (UFD) galaxies, with $M_{\ast}<10^5\msun$, tend to be devoid of gas within their virial volume, while more massive dwarf galaxies in the field often turn out to be gas-rich (e.g. \citealp{McConnachie2012}). This indicates that in order for dwarfs to be seen as DLAs, their host halo should be massive enough to shield the gas from the external UV background and to retain the gas subjected to internal stellar feedback such as supernovae. \citet{Jeon2017} also found a mass threshold of $M_{\rm vir}\gtrsim4\times10^9\msun (z=0)$ in order for a dwarf to retain gas in neutral phase until $z=0$. This study explores one of these systems in detail. At the same time, they should be in isolation, to not lose gas through environmental effects, such as ram pressure stripping and tidal interactions.

This paper is organized as follows. In Section 2, we briefly describe the simulations and adopted methodology. The main results concerning the spectra and corresponding abundance analysis are given in Section~3. We compare our results with observations in Section~4. Finally, we summarize our main findings, and discuss overall implications, in Section~5. For consistency, all distances are expressed in physical (proper) units unless noted otherwise.

\begin{figure}
  \includegraphics[width=80mm]{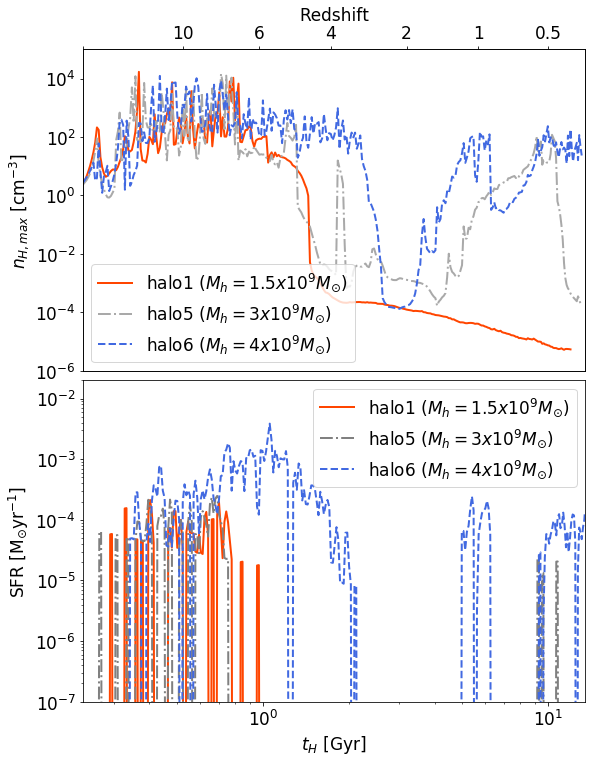}
   \caption{Assembly history of three dwarf galaxies, taken from our previous work (\citealp{Jeon2017}). The halo mass of each dwarf at $z=0$ is listed in the legend. {\it Top:} Evolution of maximum gas volume density. {\it Bottom:} Star formation rate (SFR) vs. cosmic time or redshift. From our simulations, we select the least massive system ({\sc Halo1}), and two more massive ones ({\sc Halo5} and {\sc Halo6}). Unlike {\sc Halo1}, the other two systems, {\sc Halo5} and {\sc Halo6}, have sustained gas at high density at late epochs, but no neutral H~I gas with high column density, eligible for DLAs, is found in {\sc Halo5}. Therefore the most massive halo, {\sc Halo6} is a suitable candidate to probe the DLA features associated with a small dwarf galaxy. Thus, throughout the paper ``the simulated galaxy'' refers to {\sc Halo6}.}
\end{figure}

\label{Sec:Intro}

\section{Methodology}
\subsection{Dwarf Galaxy Simulations}
We use the results of cosmological hydrodynamic simulations of dwarf galaxy formation presented in \citet{Jeon2017}, where we have performed zoom-in simulations for six sets of dwarf galaxies by tracing their evolution from the era of the first stars down to $z=0$. Here, we focus on the most massive system in our suite of simulations, {\sc Halo6}, with a mass of $M_{\rm vir}=4\times10^9 \msun (z=0)$. This is because {\sc Halo6} is the only galaxy that has maintained neutral hydrogen gas until the present-day, while the less massive systems in the mass range of $1.5\times10^9 \msun \lesssim M_{\rm vir}<4\times10^9 \msun (z=0)$ lack any dense neutral gas, which has been, predominantly, evaporated by the photo-heating during reionization at $z\gtrsim6$. We employ GADGET (\citealp{Springel2001, Springel2005}), a smoothed particle hydrodynamics (SPH) code, with well-tested modules for baryon physics implemented, specifically for star formation, population synthesis, as well as stellar radiative and supernova feedback, mainly taken and modified from the OWLS simulations (\citealp{Schaye2010}). We refer the reader to our previous work for further details (\citealp{Jeon2017}). A novelty of the simulations is that we have explicitly included the enrichment from metal-free stars, tracing the metal species released by Pop~III stars. We can thus study the imprint of the Pop~III yield signature in the second generation of Pop~II stars. 
 
Fig.~1 shows the maximum hydrogen number density (top panel) and the star formation rate (bottom panel) for three of our simulated galaxies, {\sc Halo1} with $M_{\rm vir}=1.5\times10^9 \msun (z=0)$, {\sc Halo5} with $M_{\rm vir}=3\times10^9 \msun (z=0)$, and {\sc Halo6} with $M_{\rm vir}=4\times10^9 \msun (z=0)$, as a function of cosmic time. We have introduced reionization as a uniform UV background at $z=7$ (\citealp{Haardt2012}), starting from zero strength, and increasing it to its full amplitude by $z=6$. As is evident in Fig.~1, the star formation in {\sc Halo1} has been completely shut off below $z\sim6$, whereas {\sc Halo6} has continued forming stars down to $z\sim3$, followed by bursty star formation until the present day. At $z=0$, we find stellar masses of $M_{\ast}=4.3\times 10^4\msun$ and $M_{\rm \ast}=8.8\times 10^5\msun$ for {\sc Halo1} and {\sc Halo6}, respectively. We note that {\sc Halo1} is representative of the other modeled halos ({\sc Halo2-4}) at this mass scale.  

We should note that all stars in {\sc Halo1} formed prior to reionzation, while {\sc Halo6} has continuously built up stellar mass over cosmic time. The reason for this dichotomy is that the progenitor of {\sc Halo6} was sufficiently massive to prevent complete photo-evaporation during reionization, resulting in the survival of high density gas. As a result, until $z=3$ the maximum gas density in {\sc Halo6} remains close to $\sim$ 10 $\rm cm^{-3}$, which is high enough to continue star formation. Afterwards, {\sc Halo6}  experiences episodic star formation due to the interplay between the fallback of ejected gas, mixed in with the accretion of pristine IGM material, and outflows triggered by stellar feedback. The other massive halo, {\sc Halo5}, also shows a similar SFH trend to {\sc Halo6}, as displayed as gray dot-dashed line in Fig.~1, but no neutral gas with sufficiently high column density ($N_{\rm HI}\gtrsim10^{20.3} \rm cm^{-2}$) for DLAs is found in {\sc Halo5}. We find that the fraction of time that {\sc Halo6} spends in a DLA state is $\sim35$\% over the cosmic time. The persistence of neutral gas renders {\sc Halo6} the only suitable candidate to exhibit DLA features during its assembly history. Thus, throughout this paper, ``the simulated galaxy'' specifically refers to {\sc Halo6}. We caution that the results in this work should be considered as an illustrative, individual case, rather than establishing a generalized trend, given that we pre-select a galaxy from our simulation suite which can harbor DLAs. Although, of the 6 simulated ultra-faint dwarf galaxy analogs in \citet{Jeon2017}, only one was able to maintain neutral gas at densities sufficient to produce observable DLA features. This suggests that this scenario may be rare at these mass scales.

In these simulations we self-consistently follow the evolution of heavy elements in the ISM of dwarf galaxies and the diffuse IGM, contaminated both by Pop~III and Pop~II stars. At each timestep, the masses of individual metal species newly produced in Pop~III and Pop~II stars are calculated, based on our population synthesis model. Initially, the SN ejecta are evenly distributed among the neighboring SPH particles, $N_{\rm neigh}=48$, and are subsequently transported further onto the surrounding particles. Metal diffusion is implemented by following the methodology of \citet{Greif2009}, where the diffusion coefficient of each SPH particle is computed based on the gas properties on the kernel scale. 

\begin{figure*}
    \centering
    \subfloat{{\includegraphics[width=6.6cm]{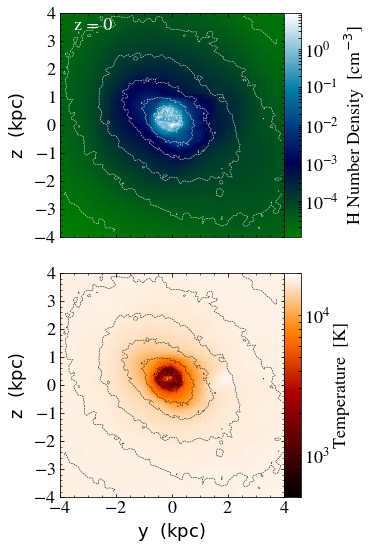} }}%
    \subfloat{{\includegraphics[width=12.04cm]{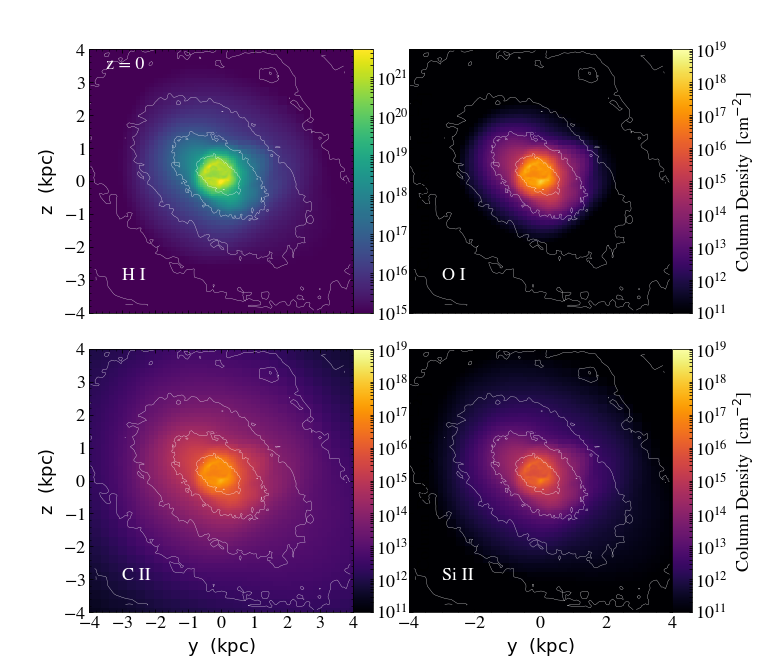} }}%
    \caption{Projection of hydrogen number density ({\it top-left}) and gas temperature ({\it bottom-left}) along the $x$-axis, centered on the simulated galaxy with $M_{\rm vir}=4\times10^9\msun$ at $z=0$, in a box with a linear size of 8 kpc. Compared to the entire extent of the dwarf galaxy, which has a virial radius of $\sim$32 kpc, high-density gas ($n_{\rm H}\gtrsim$ 1 $\rm cm^{-3}$) is only present within $\sim$1 kpc. Except in the central region, where the gas is self-shielded from external UV radiation, the majority of the gas in the dwarf is heated to above $T\sim10^4$\,K. {\it Middle and right panels}: Column densities of H~I, O~I, Si~II, and C~II (clockwise from upper-middle), evaluated at $z=0$. The column density reaches up to $N_{\rm HI}\sim10^{21} \rm cm^{-2}$ for neutral hydrogen, while the maximum column densities for metal ion species is $\sim10^{17} \rm cm^{-2}$. We also superimpose dark matter iso-density contours onto all panels, with intervals corresponding to a factor of 10 difference in dark matter density.}%
    \label{fig:example}%
\end{figure*}

\begin{figure*}
    \centering
    \subfloat{{\includegraphics[width=6.6cm]{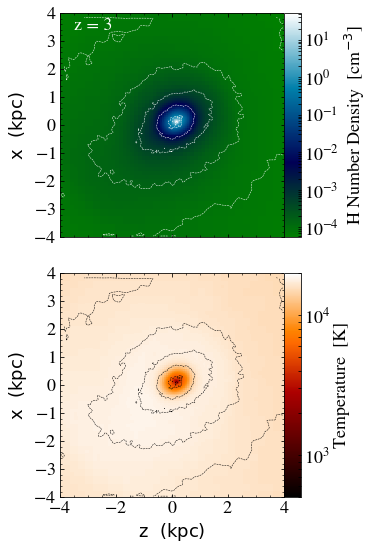} }}%
    \subfloat{{\includegraphics[width=12.04cm]{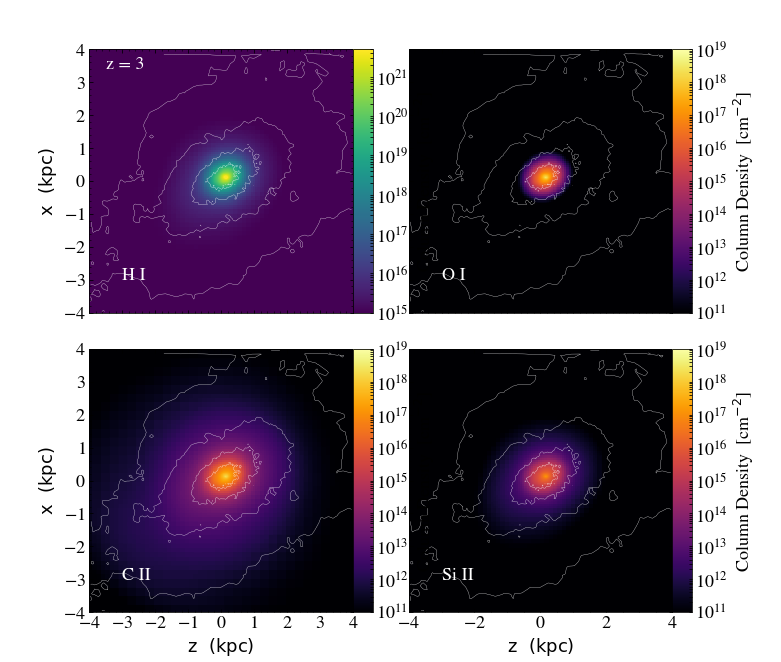} }}%
    \caption{Gas properties and column densities at intermediate redshifts ($z=3$), to be compared to Fig.~2 (for $z=0$). Here, the projections are along the $y$-axis. The corresponding virial mass and radius are $M_{\rm vir}\approx 7\times 10^8 \msun$ and $r_{\rm vir}\approx6.8$\,kpc, respectively. While the total gas mass is comparable, $M_{\rm gas}\approx2\times10^7\msun$, at both $z=3$ and $z=0$, the ISM of the galaxy at $z=3$ is more thoroughly ionized, resulting in a more compact morphology of the high density gas ($n_{\rm H}\gtrsim$ 1$\rm cm^{-3}$), which is confined to the very centre of the galaxy.}%
    \label{fig:example}%
\end{figure*}

For the metal abundance ratios of Pop~III SNe, we have adopted the IMF-weighted yields by integrating over a flat IMF, $dN/dM \propto M^{-1+\gamma}$ 
(where $\gamma=1$ for a flat IMF), in the range $10-150\msun$. Massive metal-free stars with main-sequence masses in the range $10-100\msun$ are expected to explode as CCSNe, whereas for $140-260\msun$ they are expected to die as PISNe. We adopt the Pop~III metal yields provided by \citet{Heger2002} for PISNe and \citet{Heger2010} for CCSNe. In particular, \citet{Heger2010} provide yields for Pop~III CCSNe for 120 different masses in the range of $10-100\msun$. Each star experienced an explosion, triggered by a piston located at the base of the oxygen shell, thus providing the final kinetic energy to the ejecta. We choose the final set of yields for metal-free stars with a typical explosion energy 
of $1.2\times10^{51}$ ergs and a standard mixing prescription. For Pop~II, we calculate the IMF-weighted yields by assuming a Salpeter-like IMF, employing a power-law slope of $\gamma=1.35$, together with the nucleosynthetic yields from metal-enriched stars for different metallicity, as provided by \citet{Portinari1998}.

\subsection{Column density and metal absorption spectra}
In order to generate artificial DLA spectra and the associated metal absorption lines from the cosmological hydrodynamics simulations, we use the {\sc Fake Spectra} code, developed by \citet{Bird2015}. Note that we have verified the robustness of the derived absorption spectra by additionally employing the {\sc Trident} code (\citealp{Hummels2017}), but focus here on the results from {\sc Fake Spectra}.
In general, gas column densities are computed by projecting the 3-dimensional density field onto a 2-dimensional grid along the projection axis. In order to resolve the simulated dwarfs, we assign each grid cell a linear size of $10$\,pc (comoving). We test the results by changing the cell size and verify that the derived column densities are converged at the 2\% level. Specifically, the column density, when projected along the {\it x} axis, is derived by 
\begin{equation}
N_{\rm HI} = (1+z)^2 \int{n_{\rm HI}(x) dx},
\end{equation}
where $n_{\rm HI}$ is the neutral hydrogen number density in comoving units. The ionization states of H and He are provided as outputs from the simulations, where we have solved a non-equilibrium chemistry solver at every timestep. On the other hand, the ion fractions of the metal species are computed by assuming ionization equilibrium to circumvent the computational cost of solving the corresponding network of rate equations. As for neutral hydrogen, we calculate the column density of ionization state $i$ of metal species $X$, $N_{X,i}$, by projecting its number density onto the 2-dimensional grid. In doing so, we extract the total number density of a given metal element from our simulations, where the gas metallicity is self-consistently assigned through the diffusion of metals ejected from SNe, as described in Section~2.1. Subsequently, metal ionization fractions are computed with the photo-ionization code CLOUDY (\citealp{Ferland2013}), where an interpolation is carried out over gas temperature and hydrogen number density in the presence of an external UV radiation field. For simplicity, we here assume that within a given gas cell along the projection column, gas temperatures and densities are constant. The resulting column density of a given metal ion is then given by
\begin{equation}
N_{X, i} = (1+z)^2 \int{n_{\rm X}(x) f_{X_i} dx},
\end{equation}
where $n_{\rm X}$ is the total number density of the element and $f_{X_i}$ the fraction of the $i$th ionization state, computed within CLOUDY under the assumption of ionization equilibrium. 

Here, we briefly summarize the methodology of generating synthetic absorption lines, described in \citet{Bird2015}. We choose sight-lines to penetrate the centre of a simulated galaxy, so that the highest-density neutral gas always lies in the line of sight, possibly generating a DLA system. The total absorption along each sight-line is the sum of the absorption arising from each gas cell, computed by convolving the Voigt profile, $\mathcal{V}(v)$ with a density kernel. The optical depth is computed as follows,
\begin{equation}
\tau = \mathcal{V}(v) \times n\left( \sqrt{v^2/H^2(z) + r^{2}_{\rm perp}} \right), 
\end{equation}
where a redshift space position, $v$, is defined as $v=H(z) x + v_{\rm par}$. Here, $H(z)$ is the Hubble parameter at a given redshift, $x$ the position of the gas cell in the box, and $v_{\rm par}$ the velocity of the cell parallel to the line of sight. The physical quantities are interpolated along a sight-line using the smoothing kernel, $n\left( \sqrt{v^2/H^2(z) + r^{2}_{\rm perp}} \right)$, which is a function of the distance from the cell centre to a point in the cell, where $r_{\rm perp}$ is the distance perpendicular to the line of sight. The smoothing length is chosen to be the radius of a sphere comparable to the volume of the grid cell. We choose the simulated spectra to have a spectral pixel size of 0.01 km $\rm s^{-1}$.



\section{Results}
In the following, we present our results. First, in Section 3.1, we
present the properties of DLAs associated with the simulated galaxy. In particular, we 
discuss the distribution of column densities of the gas and present the resulting 
absorption spectra. Based on the physical properties of the DLAs, we discuss the derived relation between velocity width and metallicity in the galaxy. In Section 3.2, we analyze 
the abundances patterns of the DLAs and compare with those of observed dwarfs. 

\begin{figure}
  \includegraphics[width=80mm]{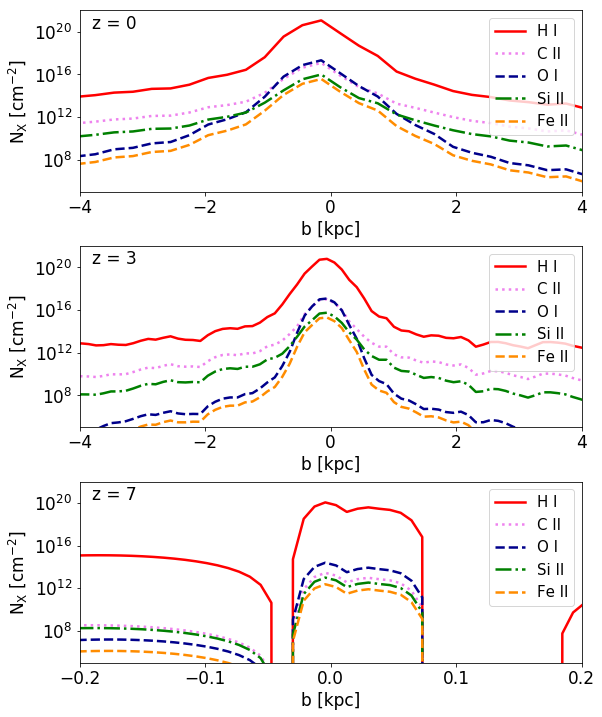}
   \caption{Column density of H\,I and metal ion species at various impact parameter, $b$, along the $x$-axis at $z=0, 3,$ and $7$ ({\it top to bottom}). The high column density gas that satisfies the DLA threshold, $N_{\rm HI}>10^{20.3} \rm cm^{-2}$, is located near the galactic centre, $r\lesssim$500\,pc at $z=0$, while the gas is more concentrated at $z=3$. For the minihalo progenitor at $z=7$, the virial radius is $\sim100-200$\,pc, and we therefore show gas column densities only within 200\,pc ({\it bottom panel}).}
\end{figure}

\begin{figure}
  \includegraphics[width=80mm]{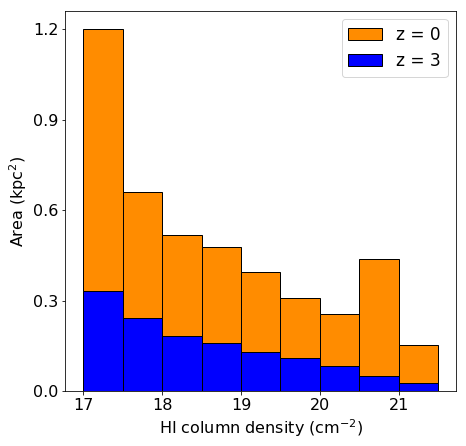}
   \caption{The cross-section, or area occupied by neutral gas in the simulated dwarf at $z=0$ (orange) and $z=3$ (blue), as a function of $N_{\rm H\,I}$ column density. The projected area for gas with column density larger than $N_{\rm H\,I}=10^{20.3}\rm cm^{-2}$ is about $\approx$1 kpc$^2$ at $z=0$, and it decreases to $\approx$0.2 kpc$^{2}$ at $z=3$. We interpret this difference between $z\approx3$ and $z\approx0$ as a consequence of the variation in the ionization state due to stellar feedback. Also, the bump at $N_{\rm H~I}\sim10^{21}\rm cm^{-2}$ ($z=0$) is a particular feature of the simulated galaxy. We emphasize that although the cross-sections in dwarf galaxies are small, the fact that low-mass galaxies are more abundant, owing to the steep faint-end slope of the luminosity function, could compensate in setting the overall probability of detecting DLAs associated with small dwarf galaxies (e.g., \citealp{Krogager2017}).}
\end{figure}

\subsection{DLAs in the simulated galaxy}
We focus on the gas properties within the simulated dwarf galaxy, 
at three representative redshifts, corresponding to the early evolution at $z=7$, an intermediate snapshot at $z=3$, and the present-day situation at $z=0$. With those properties in hand, we are then able to construct synthetic spectra for the possible DLA phases.

\subsubsection{Column densities}
Fig.~2 shows the hydrogen number density (top-left panel) and the gas temperature (bottom-left panels) in a box with a linear size of 8 kpc, centered on the simulated halo at $z=0$. While its virial radius is $r_{\rm vir}\sim$32 kpc, the high density gas, 
$n_{\rm H}\gtrsim1$ $\rm cm^{-3}$, only exists in the central region within $\sim1$ kpc. As seen in the temperature map, a majority of gas, approximately $70\%$, is heated to above $T\sim10^4$\,K except for the self-shielded gas at the centre, where the average temperature is about $10^{3.5}$\,K. We note that the gas within $r\lesssim50$\,pc from the centre is pushed out, establishing a ring-like structure, as can be seen in the figure. This is owing to the ongoing star formation, followed by SN energy input, with an estimated star formation rate of about $\dot{M}_{\ast}=10^{-4}\msun \rm yr^{-1}$ at $z=0$ (see bottom panel of Fig.~1).

Unlike observed UFDs in the vicinity of the Milky Way, the simulated dwarf shows a substantial amount of gas in the neutral phase. Specifically, we have found a mass of warm neutral hydrogen of $M_{\rm H\,I, w}=2.8\times10^6\msun$, corresponding to $10\%$ of the total gas, and $M_{\rm H\,I, c}=9.3\times10^4\msun$ in a cold H\,I medium. Here, we define the warm and cold H\,I phases according to $T\sim10^4$\,K and $T\lesssim300$\,K, respectively (e.g., \citealp{Wolfire2003}). We note that we have targeted an isolated dwarf galaxy, indicating that certain environmental effects, such as ram pressure stripping and tidal interactions, are not present. Otherwise, any remaining neutral gas could have been stripped away by these effects. Therefore, we may consider our simulated galaxy as an analog of local field dwarfs, such as Leo~P or Leo~T, which are gas-rich dwarfs with a neutral hydrogen gas content of $M_{\rm H\,I}=2.8\times10^5\msun$ and $M_{\rm H\,I}=9.3\times10^6\msun$, respectively.

In Fig.~2, we also present column densities for neutral hydrogen and select metal species, specifically neutral oxygen, singly-ionized silicon, and singly-ionized carbon (middle and rightmost column). As can be seen, the gas with high H\,I column density, $N_{\rm H\,I}\gtrsim10^{20.3}$ $\rm cm^{-2}$, eligible to produce DLA systems, is concentrated within the central $r\lesssim1$\,kpc. This implies that in order to observe DLA systems, arising from neutral gas within a dwarf galaxy, lines of sight need to penetrate its central region. To illustrate the evolution of the system, in Fig.~3 we show the same quantities as in Fig.~2, but for $z=3$. At this earlier stage, the virial mass of the simulated galaxy is $M_{\rm vir}\approx 7\times 10^8 \msun$, and the virial radius $r_{\rm vir}\approx6.8$\,kpc. 


In Fig.~4, we compare column densities at a different impact parameter $b$ for three illustrative redshifts. At $z=0$, the H\,I column density rises to $N_{\rm H\,I}\approx10^{21}\rm cm^{-2}$ at the centre ({\bf $b=0$}), and gradually drops with increasing impact parameter, reaching $N_{\rm H\,I}\approx10^{14}\rm cm^{-2}$ at 4\,kpc. Column densities of singly-ionized carbon and silicon are broadly distributed, as well, with central values of $N_{\rm C\,II}\approx10^{16} \rm cm^{-2}$ and $N_{\rm Si\,II}\approx10^{15} \rm cm^{-2}$, respectively. On the other hand, O\,I and Fe\,II column densities more steeply decrease with increasing distance from the center. This is because oxygen and iron are likely to be in their singly- (O\,II) and doubly-ionized (Fe\,III) states beyond $r\sim1.2$\,kpc, where the diffuse ISM, $n_{\rm H}<10^{-3}\rm cm^{-3}$, is heated to $T \gtrsim10^4$\,K. As shown in the middle panel of Fig.~4, at $z=3$ 
high column density gas is more concentrated. In particular, H\,I column density decreases from $N_{\rm H\,I}\approx3\times10^{20}\rm cm^{-2}$ by four orders of magnitude at 700\,pc, while $N_{\rm HI}\approx10^{16}\rm cm^{-2}$ is reached at a radius of 1.6\,kpc at $z=0$. For $z=3$, O\,I and Fe\,II column densities also drop more sharply.

At $z=7$, we show column density distributions within a minihalo, which is a progenitor in the assembly tree of the dwarf, with a virial mass and radius of $M_{\rm vir}=10^6\msun$ and $r_{\rm vir}=200$\,pc. Therefore, we present column densities only within 200\,pc in the bottom panel of Fig.~4. 
The central high column density gas is fairly evenly distributed over a radius of 100\,pc, and sharply truncated beyond. We note that the H\,I column density is as high as $N_{\rm H\,I}\sim10^{20}\rm cm^{-2}$ in the minihalo, marginally sufficient to imprint a DLA feature. If the gas in a minihalo is only contaminated by Pop~III stars, which is plausible in such low-mass settings, it would provide us with a unique opportunity to probe the first generation of stars through such DLA observations \citep[e.g.][]{Jaacks2018a}.

As is evident in Figs.~2 and 3, the high column density gas, able to imprint a DLA feature, is only located at the centre of the galaxy. In order to study the concentration of high column density gas, we present cross-sections, the areas occupied by a given column density, in Fig.~5. Specifically, the area occupied by gas with column density larger than $N_{\rm HI}=10^{20.3} \rm cm^{-2}$, which is the definition of a DLA system, is about $\approx$1 $\rm kpc^{2}$ at $z=0$, reduced to $\approx0.2 {\rm kpc}^{2}$ at $z=3$. We found that the total amount of gas in the dwarf is roughly the same, at $M_{\rm gas}\approx4\times10^7\msun$ both at $z=3$ and $z=0$, whereas the halo virial mass grows from $M_{\rm vir}\approx7\times10^8\msun$ ($z=3$) to $M_{\rm vir}\approx4\times10^9\msun$ ($z=0$). This is due to the vulnerability of the dwarf caused by stellar feedback. Immediately after $z\approx3$, the gas is significantly heated and ionized both by stellar feedback from in-situ stars and external UV radiation (see Fig.~1 and also Fig.~2 in \citet{Jeon2017}), lowering the total amount of gas mass by a factor of $\sim5$ by $z\approx2.3$. Afterwards, the gas is replenished through mergers, allowing the galaxy to recover the total amount of gas with $M_{\rm gas}\approx4-5\times10^7\msun$ at $z\lesssim2$. Given the similar amount of total gas in the galaxy, the difference in the cross-section between $z=3$ and $z=0$ can be attributed to the ionization state of the gas. This is clearly demonstrated in Figs.~2 and 3, showing that the galaxy's ISM is significantly more ionized earlier on, with a fraction of warm neutral gas of $\approx4\%$ at $z=3$, while it is $\approx20\%$ at $z=0$. We emphasize, however, that such comparison derived in this work is based on a single dwarf galaxy, thus not representing a general trend.

As such, the probability of generating DLA systems associated with dwarf galaxies ($M_{\rm vir}\lesssim10^9\msun$) might be small, since the photon path from a background source ought to penetrate the central region of the dwarf galaxy. At $z\sim3$, the area occupied by high H\,I column density further decreases as the ISM is ionized by stellar feedback. We should note that even though the neutral gas cross-sections of the progenitors of low-mass dwarfs ($M_{\rm vir}\approx4\times10^9\msun (z=0)$) is small, the effect of luminosity function weighting, where small galaxies are more abundant, could boost DLA detections associated with such tiny galaxies (e.g., \citealp{Fynbo1999}). Recent results by \citet{Krogager2017} have suggested that the DLA observational data suggests a very flat probability distribution across luminosities by weighting with the luminosity function. 

\begin{figure}
  \includegraphics[width=90mm]{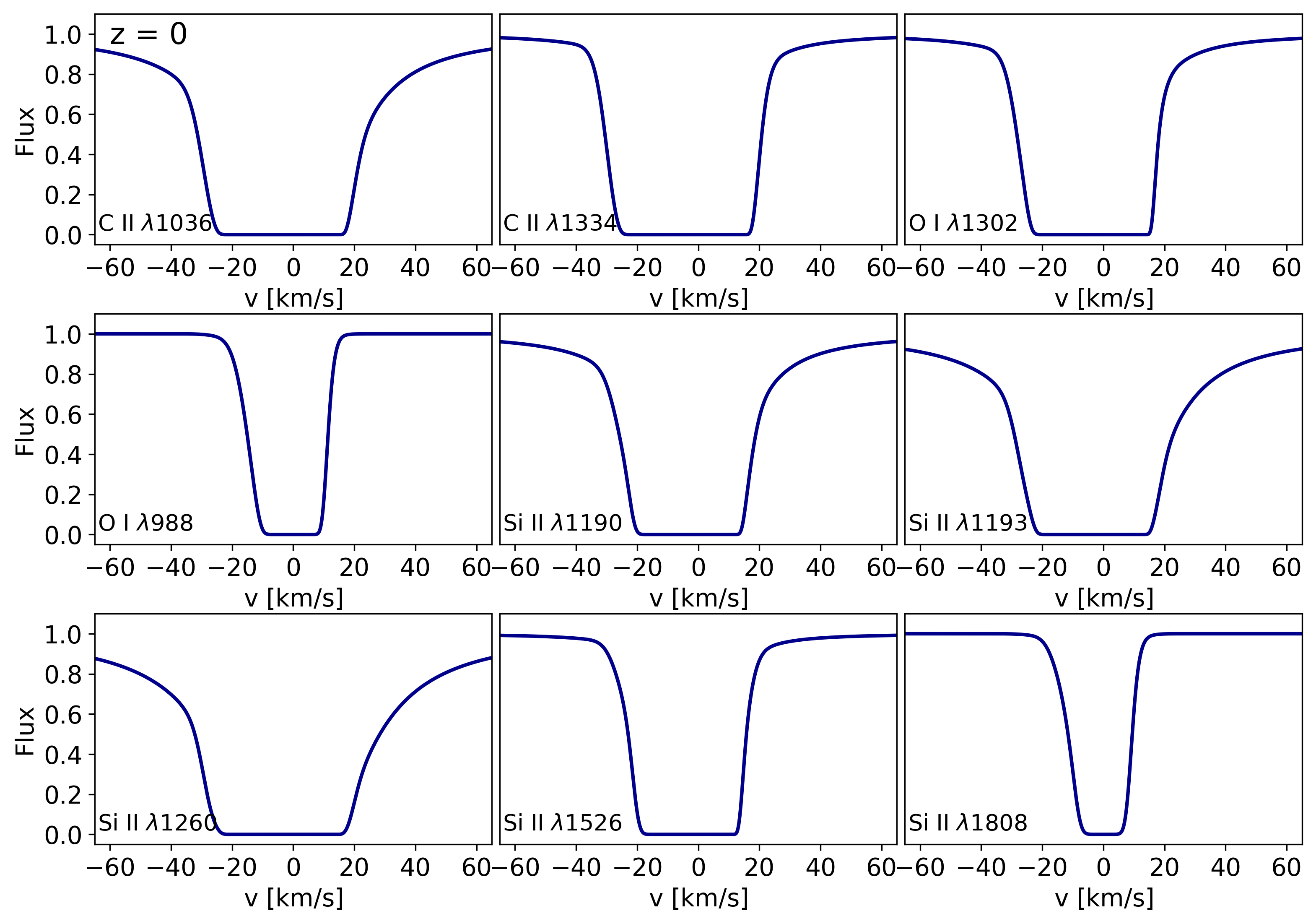}
   \caption{Select metal absorption lines associated with the DLA gas at $z=0$. The C\,II, O\,I, and Si\,II lines are all saturated, owing to the high gas metallicity of $Z\approx0.12\zsun$.}
\end{figure}

\begin{figure}
  \includegraphics[width=90mm]{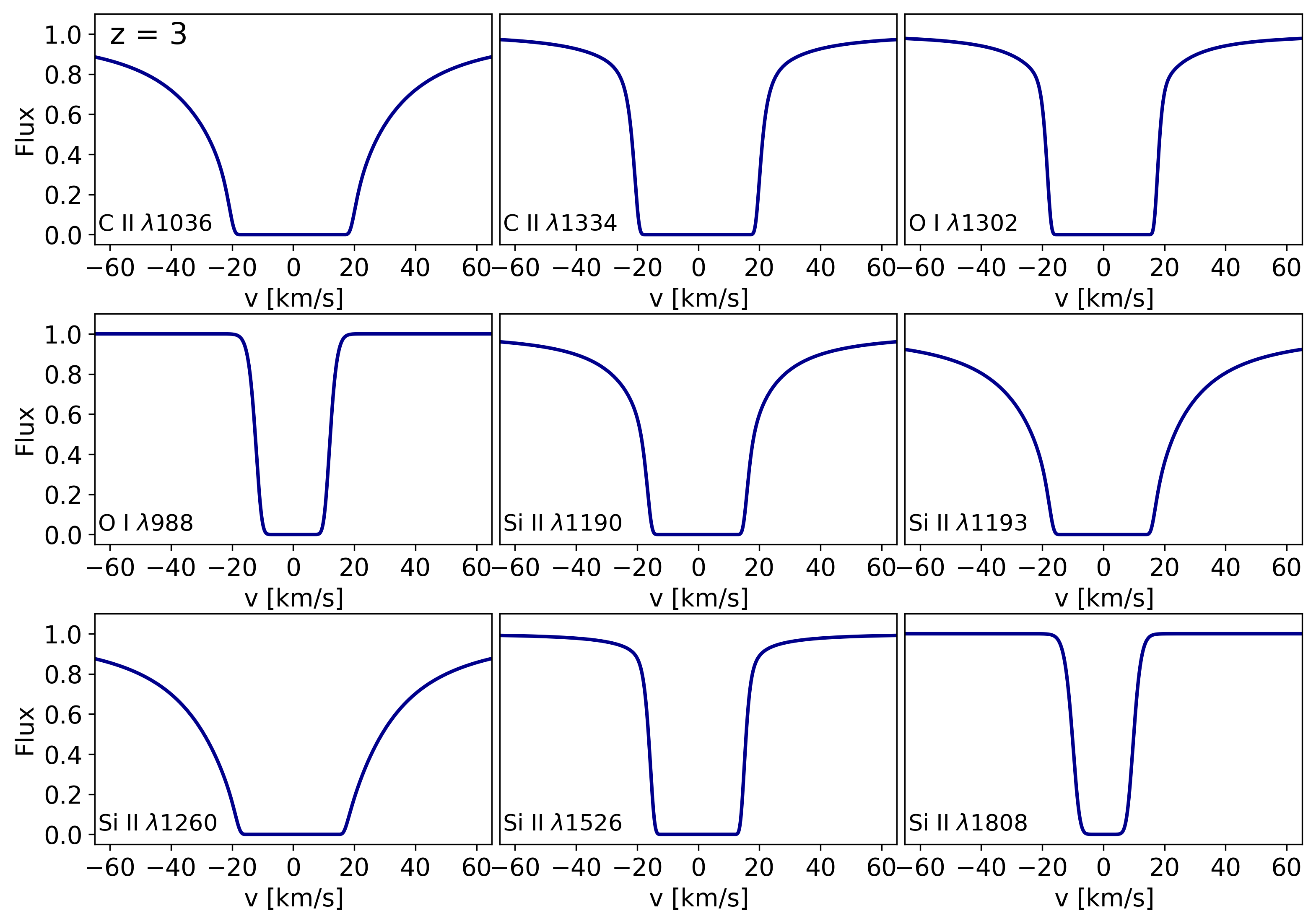}
   \caption{The same as Fig.~6, but for $z=3$. Again, all lines are strong enough to achieve saturation.}
\end{figure}

\begin{figure}
  \includegraphics[width=90mm]{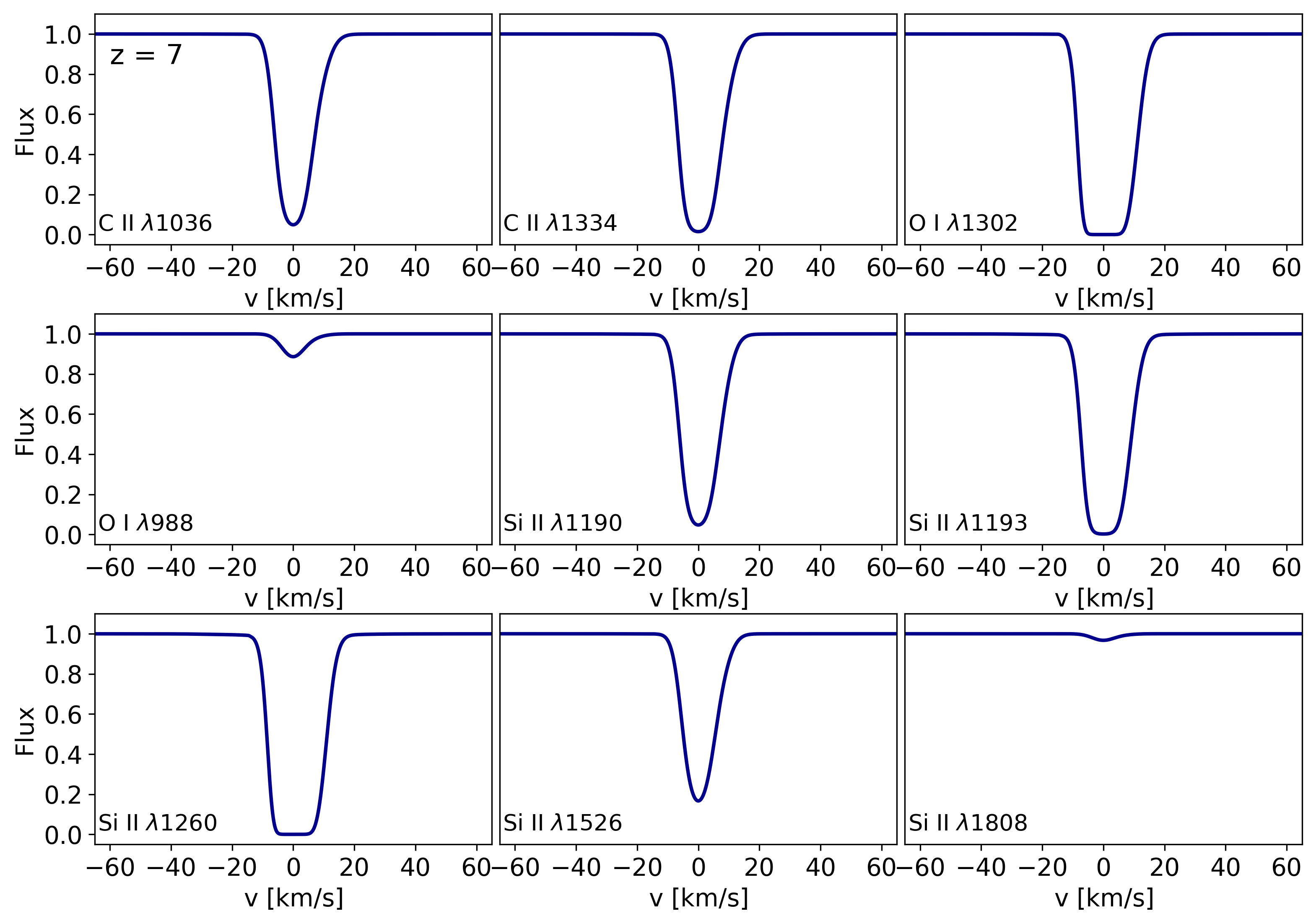}
   \caption{The same as Fig.~6, but for $z=7$. At this early time, most lines are still weak, reflected in the unsaturated line profiles.}
\end{figure}

\begin{figure}
  \includegraphics[width=90mm]{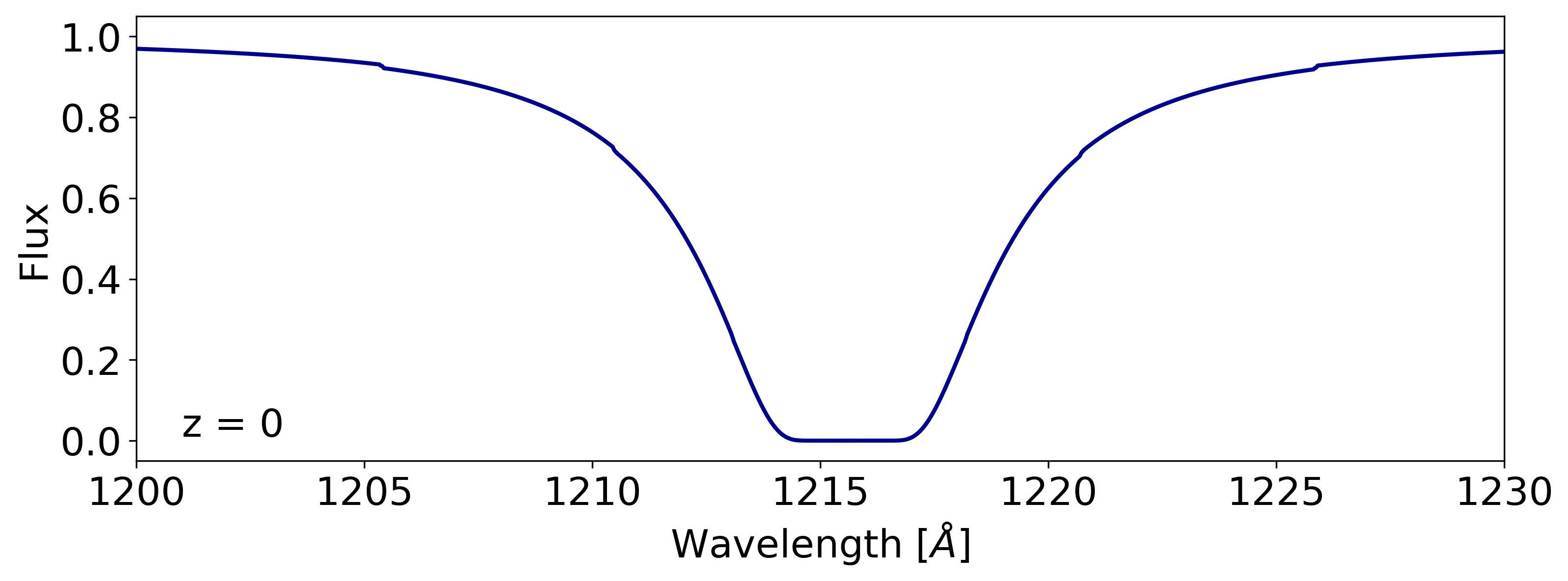}
   \caption{Damped Ly$\alpha$ absorption profile, imprinted by the neutral gas in the centre of the dwarf galaxy at $z=0$.}
\end{figure}

Aside from the main target galaxy, {\sc Halo6}, we also consider an additional simulated dwarf, {\sc Halo5}, that reaches a virial mass of $M_{\rm vir}\approx3\times10^9\msun$ $(z=0)$, experiencing brief episodes of star formation at intermediate redshifts, for example at $z=0.5$ (see Fig.~1). In order to ascertain if this dwarf might produce a DLA system, we compute the H\,I column density in {\sc Halo5} at the moment of star formation. Interestingly, there is no noticeable neutral gas with column density as high as $N_{\rm HI}\gtrsim2\times10^{20}\rm cm^{-2}$. This implies that the observability of neutral gas, in the form of DLAs, will also be sensitive to the timing of the observations, since neutral gas in dwarfs is likely to be ionized and dispersed by stellar feedback. Thus, the most promising moment to observe a DLA phase would be immediately prior to a star formation event. Interestingly, \citet{Moller2018} have reported a DLA-selected galaxy at $z=0.716$, which is the second-most massive DLA-selected galaxy known, with stellar mass of $\approx 10^{10.8}\msun$. The DLA has a large molecular gas reservoir of $M_{\rm mol}=10^{10.4}\msun$, but has no counterpart in flux-limited samples. The authors have suggested that such failure to detect a counterpart could imply that the galaxy might be in between a starburst and post-starburst phase, resulting in a wide CO profile and low star formation rate. Additional recent observations of DLA and sub-DLA galaxies with relatively low SFRs (\citealp{Kanekar2018, Rhodin2018}) have supported a post-starburst scenario. Molecular gas densities would then be reduced by the feedback from previous star formation, and an extended molecular disk would render the galaxy more easily detectable as DLAs.

\subsubsection{Synthetic absorption spectra}
A selection of absorption lines imprinted by different metal species is shown for the simulated dwarf galaxy at $z=0$, $3$, and $7$ in Figs.~6, 7, and 8, respectively. As explained in Section~2, we employ the {\sc fake spectra} package, which generates absorption profiles along a single ray. We choose a sightline that passes through the dense neutral cloud at the centre of the simulated galaxy. Specifically, we investigate select heavy-element absorption features for low-ionization species, such as C\,II, Si\,II, and O\,I, which are known as suitable tracers of the kinematic and density structure in DLA systems (e.g. \citealp{Prochaska1997}). In Figs.~6-8, we show spectra in velocity space with a unit of km~$\rm s^{-1}$ for the simulated galaxy at different redshifts. Finally, we also show the damped Ly$\alpha$ profile for the $z=0$ case in Fig.~9.

At $z=0$, the resulting metal absorption lines arising from the gas at the centre of the simulated galaxy exhibit saturated absorption features owing to the high column densities present there. As shown in Fig.~4, we find $N_{\rm C\,II}\approx10^{16} \rm cm^{-2}$, 
$N_{\rm O\,I}\approx10^{16} \rm cm^{-2}$, and $N_{\rm Si\,II}\approx10^{15} \rm cm^{-2}$, respectively. At $z=3$, gas column densities for the metal species are similarly high, again resulting in saturated line shapes. On the other hand, while the H\,I column density in the minihalo at $z=7$ can reach as high as $N_{\rm H\,I}\approx10^{20} \rm cm^{-2}$, the column densities associated with the metal ions are lower by 5-6 orders of magnitude, leading to weak, unsaturated lines. The reason is that the gas in the minihalo is barely enriched, 
with typical metallicities of $\lesssim 10^{-3}\zsun$ at $z=7$. We note that the DLA features are produced by penetrating the gas cloud along a given axis, and the resulting absorption lines are likely to show nearly symmetric profiles in velocity space. However, the nature of the absorption features can change, depending on the density and velocity structure of the gas. 

\begin{figure}
  \includegraphics[width=90mm]{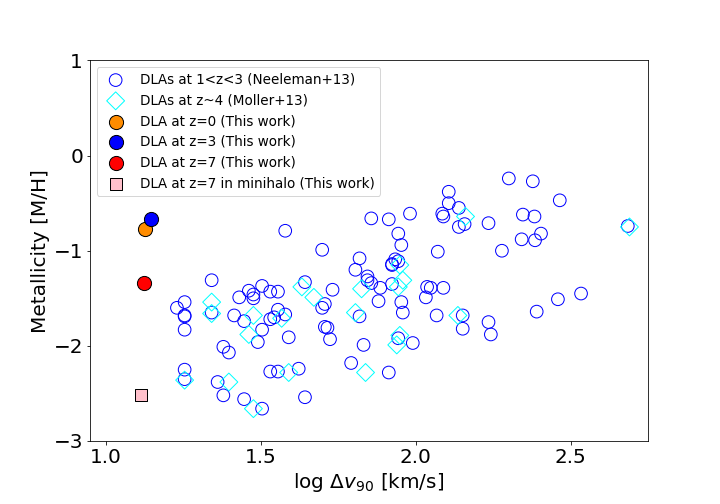}
   \caption{Metallicity vs. velocity width in DLAs. The observations, denoted by cyan diamonds (\citealp{Moller2013}) and blue open circles (\citealp{Neeleman2013}), indicate that DLA metallicities become larger with increasing $\Delta v_{90}$, which is a proxy for stellar mass. For $z=7$, we show two DLAs located in the main halo, the most massive progenitor halo (red filled circle), and in a minihalo (pink square), where a Pop~III DLA resides, exhibiting unique signatures of Pop~III stars. As estimated for the simulated dwarf, $\Delta v_{90}$ remains small across cosmic time. We find that $\sim$70\% of stars have already formed at $z=3$, giving rise to a stellar mass of $M_{\ast}={6.1\times10^6\msun}$. The estimated metallicities of the DLAs in this work are higher than the observed values. One of the possible reasons for this is that the DLAs are associated with central star forming gas cloud which substantially metal-polluted, leading to an offset from the observed metallicity-$\Delta v_{90}$ trend.}
\end{figure}

\begin{figure}
  \includegraphics[width=90mm]{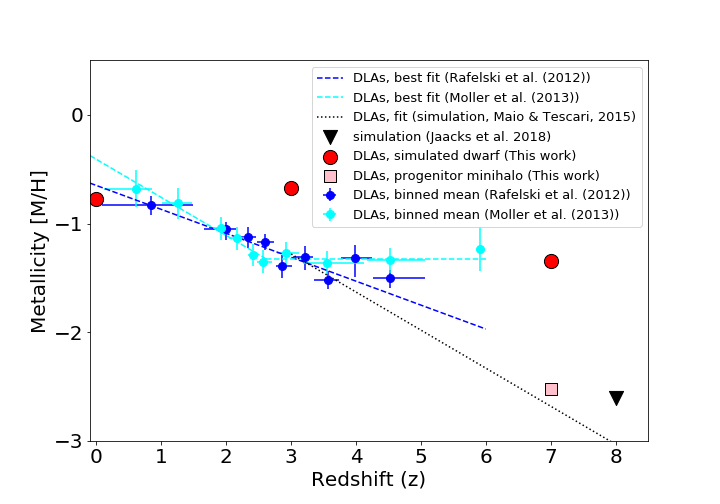}
   \caption{The gas metallicity derived from DLA observations (\citealp{Moller2013, Hartoog2015}; cyan circles, \citealp{Rafelski2012}; blue circles) as a function of redshift, in comparison with the metallicities of DLAs associated with the simulated dwarf galaxy (red filled circles and pink square). The observed values are the averaged values within a binned redshift and the corresponding best fit is shown as the cyan and blue solid line, respectively. For high redshifts ($z>6$), we superimpose the fit (black dotted line) and the estimate (black inversed triangle) that are computed from cosmological simulations for DLAs at high-$z$ Universe (\citealp{Maio2015, Jaacks2018}). The derived metallicities of the DLAs in the simulated galaxy are similar or larger than observed values for DLAs. We, however, should keep in mind that we are comparing a single simulated dwarf galaxy with the average values from observations.}
\end{figure}

\subsubsection{Velocity width and metallicity}
It has been suggested that DLAs could be a powerful tool to study the population of star-forming galaxies over a wide range of properties (e.g., \citealp{Ledoux2006, Murphy2007, Prochaska2008, Fynbo2008, Moller2013, Neeleman2013, Jorgenson2013, Christensen2014, Krogager2017}). While a mass-metallicity (MZ) relation drawn from emitting galaxies is likely to be biased towards luminous galaxies, DLAs tend to sample galaxies across the luminosity range. Using the observed velocity widths of metal absorption line profiles as a proxy for stellar mass of the galaxies, \citet{Ledoux2006} have suggested that DLAs follow an absorption-based version of the mass-metallicity relation (DLA MZ). The redshift evolution of this relation has been investigated with enlarged samples by, for example, \citet{Neeleman2013} and \citet{Moller2013}. Furthermore, \citet{Christensen2014} have confirmed the DLA MZ relation predicted by \citet{Moller2013} through direct measurements of the stellar masses of 12 DLA galaxies.

In order to verify whether this trend also applies to the present work, we compute the corresponding widths for the metal lines considered. As originally proposed, the $\Delta v_{90}$ line-width measure  corresponds to the velocity range that covers 90\% of the cumulative optical depth across a DLA, and is defined as $\Delta v_{90}=v_{\rm high} - v_{\rm low}$. Here, $v_{\rm high}$ and $v_{\rm low}$ are the velocities where the integrated optical depth is 95\% and 5\% of the total, respectively. Specifically, the Si\,II~1808\,{\AA} transition line, a typical example of an unsaturated line, is often used 
to evaluate $\Delta v_{90}$. For our simulated galaxy, we estimate $\Delta v_{90}$=13.4, 14, 13 km $\rm s^{-1}$ at $z=0$, $3$, and $7$, respectively.
The halo mass increases from $M_{\rm vir}=5\times{10}^8\msun$ at $z=3$ to $M_{\rm vir}\approx4\times10^9\msun$ at $z=0$, in which the stellar mass is 
$M_{\ast}\approx6.1\times10^5\msun$ at $z=3$, which is 70\% of the final stellar mass $M_{\ast}=8.8\times10^5\msun$ at $z=0$. At $z\approx7$, we show the properties of two DLAs, which 
reside in the main progenitor halo, i.e., the most massive halo, and in the minihalo that will eventually merge into the main progenitor halo. The reason we additionally choose the latter is that this minihalo is the only place where we find a DLA that is solely contaminated by first generation stars, while the DLA metal enrichment in the main halo has already significantly progressed by both Pop~III and Pop~II stars, thus erasing pure signatures of Pop~III stars. At $z\approx7$, the virial (stellar) masses of the main halo and the minihalo are $M_{\rm vir}\approx3\times10^8\msun$ $(M_{\ast}\approx1.9\times10^5\msun)$ and $M_{\rm vir}\approx5\times10^6\msun$ $(M_{\ast}\approx10^4\msun)$, respectively.

We find the metallicity of the DLA phase in our simulated galaxy to be $\rm [M/H]\approx-0.8$ at $z=0$, which is larger than the observed values by a dex at a given velocity width, as seen in Fig.~10. The gas, responsible for the DLAs in the simulated dwarf, is located near the centre of the galaxy, implying an impact parameter $b$ close to 0. This is also the region where star formation mainly takes place, resulting in a high metallicity of the surrounding medium, compared to the gas in the outskirts. Therefore, we expect a gradient in metallicity, such that the metallicity decreases with increasing impact parameter. Observationally, \citet{Christensen2014} suggest a metallicity offset of $\Delta \rm [M/H]\sim-0.022\pm0.004$ dex $\rm kpc^{-1}$. Our simulated galaxy shows a similar metallicity gradient, $\Delta \rm [M/H]\approx-0.026\pm0.001$ dex $\rm kpc^{-1}$ at $z=0$. 

Fig.~10 shows that the velocity with, $\Delta v_{90}$, barely changes over cosmic time, while the metallicity increases from $\rm [M/H]\approx-1.34$ at $z\approx7$ to $\rm [M/H]\approx-0.67$ at $z=3$. We note that the metallicity of the Pop~III DLA that resides in the minihalo at $z\approx7$ is $\rm [M/H]\approx-2.52$ (pink square), which is $\sim$1.2 dex lower than the DLA in the main progenitor halo. The reason that our simulation tends to show higher metallicity than the observed values is that, as shown in Section~3.3.1, the simulated DLAs are associated with the very central gas cloud ($r\lesssim500$ pc), and hence the most polluted one. The simulated galaxy, moreover, is distinctly low-mass ($M_{\ast}<10^6\msun$), unable to shield against the external radiation which ionizes the gas in the outer parts of the galaxy, thus confining the neutral gas to the central region. On the other hand, the observed DLA absorbers appear to be associated with more massive galaxies with stellar masses $M_{\ast}\gtrsim10^8\msun$ (\citealp{Christensen2014}), such that DLAs can be associated with the gas throughout a galaxy.


In Fig.~11, we compare the metallicity of the gas associated with the DLAs at $z=0$, $3$, and $7$ from this work (red filled circles and pink square) with mean DLA metallicities from observations (\citealp{Rafelski2012} (blue circles) and \citealp{Moller2013} (cyan circles). For the highest redshift, the result is derived from a GRB DLA at $z\approx5.9$ (\citealp{Hartoog2015})). We also compare with the results from cosmological simulations (\citealp{Maio2015, Jaacks2018}) in the higher redshift regime ($z>6$). The values from \citet{Moller2013} correspond to the binned metallicity from DLAs, which have a velocity width of $\Delta V=100$ km $\rm s^{-1}$, and the cyan and blue dashed lines are the best fit to the data of \citet{Moller2013} and \citet{Rafelski2012}, respectively. The redshift evolution indicates a similar trend for both data sets at redshifts lower than $z=2.6$, while the results from \citet{Moller2013} favour a rather flat (no) evolution at high redshifts $z>2.6$, which is further supported by the data from GRB DLAs (\citealp{Arabsalmani2015, Hartoog2015}). The derived metallicity of the simulated DLA at $z=7$ in the main progenitor halo also shows a similar metallicity of $\rm [M/H]=-1.34$. The derived DLA metallicity in the simulated galaxy at $z=3$, on the other hand, is larger than the values from observations. We emphasize that our simulated DLA metallicity is based on a single dwarf galaxy, while the observed metallicities are cross-section weighted averages over a range of galaxy mass. Additionally, as previously mentioned, the high metallicity of the DLAs in the simulated dwarf arises because the DLAs originate from the central star-forming gas clouds.

We note that we have ignored radiative transfer of ionizing photons. Thus, one possible numerical reason for the presence of high-metallicity gas at the centre of the galaxy is the lack of photo-ionization heating prior to a SN explosion. Although this process in itself may not be sufficient to evacuate the enriched gas from the system at late epochs, photo-ionization heating could have played an important role in transporting metal-enhanced gas to larger distances. Otherwise, the polluted gas is likely to remain at the centre of the galaxy due to the deeper potential well at late times. The importance of including photo-ionization heating is also pointed out in \citet{Simpson2013}, where the 1\,-\,2~dex difference in stellar metallicity between observations and their simulations is attributed to the inability of expelling metals to larger distances in the absence of photo-ionization heating or radiation pressure. Recently, \citet{Corlies2018} have analyzed metal enrichment histories of dwarf galaxies at $z\gtrsim7$, based on the simulations of \citet{Wise2012}, where photo-ionization heating and radiation pressure feedback from SNe were taken into account. They found more widely distributed metals further into the IGM, but stellar metallicities within the analyzed galaxies were still higher than observations, establishing the need for further improvement in the simulations.

\subsection{Abundance Analysis}
One of the important questions associated with observed DLAs is which stellar populations have contributed to their enrichment, and at what cosmic times. To understand the enrichment history of our simulated DLAs we derive the abundances of select metal species from our simulation, and compare them with those inferred from population synthesis models for Pop~III and Pop~II stars. We aim to explicitly test the intriguing scenario that metal-poor DLA systems might correspond to gas clouds that have experienced only a single episode of enrichment from first generation star formation, or a small number of such enrichment cycles.

In Fig.~12, we present estimates for select metal yields from Pop~III and Pop~II star formation (black solid and dashed lines, respectively). We compare the two population models with the metal ratios in our simulated DLA clouds, at $z=0$, $3$, and $7$. At low and intermediate redshifts, $z=0$ and $3$, the metal ratios derived from our simulations are consistent with the Pop~II enrichment pattern, as can be seen in the middle panel of Fig.~12. On the other hand, as shown in the bottom panel, the abundance ratios at $z=7$ are in good agreement with the Pop~III pattern. We conclude that the period during which gas clouds exhibit the signature of exclusively Pop~III enrichment is very brief, confined to redshifts $z\gtrsim 7$ and to gas residing in minihaloes. Such gas clouds are likely to be polluted by subsequent Pop~II star formation, erasing the memory of Pop~III metal yields.

\begin{figure}
  \includegraphics[width=80mm]{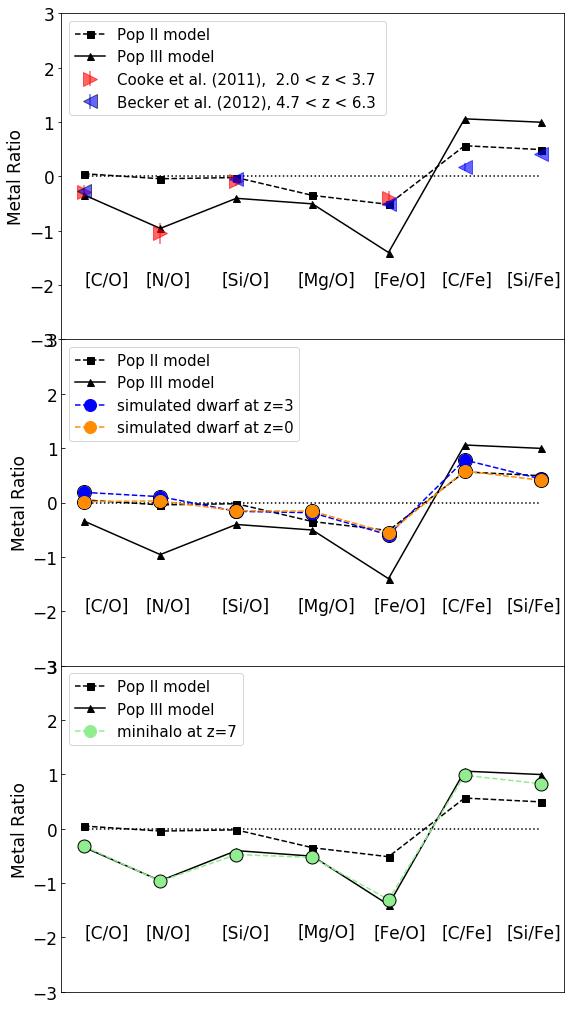}
   \caption{DLA abundance patterns. Shown are select metal ratios derived from population synthesis models for Pop~III (solid black line; triangles) and Pop II stars (dashed black line; squares), compared to the ratios associated with the DLA progenitors of the simulated dwarf galaxy and observations of real DLAs. ({\it Top panel}): Black triangles and squares connected with black solid and dashed lines represent the ratios of metal species that are produced by SNe from Pop~III progenitors in the mass range of $10-150\msun$ and from Pop~II stars with $1-100\msun$, respectively. We also display observations for metal-poor DLAs at $2.0<z<3.7$ (\citealp{Cooke2011}) with red triangles, and for higher redshift DLAs at $4.7 < z < 6.3$ (\citealp{Becker2012}), denoted by blue triangles. ({\it Middle panel}): Metal ratios associated with the DLA progenitors of the target dwarf at $z=0$ and $3$. They are in a good agreement with the metal yields synthesized by Pop~II SNe. ({\it Bottom panel}): The gas clump in a minihalo contains metal ratios which are consistent with those created by Pop~III SNe. We only find a Pop~III dominated pattern for gas inside a minihalo at $z\gtrsim 7$, since the Pop~III metal ratios are quickly wiped out by enrichment from Pop~II SNe.}
\end{figure}

The moment when the transition from Pop~III to Pop~II occurs depends on the mass of a halo that hosts Pop~III stars; smaller haloes are more susceptible to SN feedback, as they are unable to retain their gas. In addition, more energetic SNe can render a host halo increasingly sterile, leading to a longer time delay between Pop~III and Pop~II star formation. This delay can be quantified by the recovery time, when Pop~II stars can form after the initial episode of Pop~III activity. Specifically, the recovery timescale is of order a few tens of Myr for a typical minihalo that experiences a CCSN, while it can reach a few hundred Myr when a Pop~III star dies as a PISN with an explosion energy of $E_{\rm SN}=10^{52}$ ergs (e.g., \citealp{Jeon2014}).

As an extreme case, instead of achieving the Pop~III-Pop~II transition, a halo could only experience Pop~III star formation without any further second generation contribution. The gas within the halo would then have been enriched only by Pop~III SNe \citep[e.g.][]{Frebel2012}. Such conditions, however, would only be possible in low-mass minihaloes, where the gravitational potential is too shallow to re-assemble the gas evacuated by Pop~III SN feedback. Star formation would thus be permanently quenched in such a system. Furthermore, the minihalo should be isolated and not merge with other haloes, which would likely trigger further star formation as the gas supply would be replenished. 
However, the minihaloes explored in this work are progenitors of the target dwarf galaxy, thus merging into a larger central halo, rather than remaining in isolation. As displayed in Fig.~12, the occurrence of gas with a Pop~III abundance pattern is limited to minihaloes at high-$z$. Afterwards, gas in the progenitor haloes is continuously contaminated by second generation stars, thus rapidly erasing the pure Pop~III signature. Our findings are in line with results from previous studies \citep[e.g.][]{Salvadori2012, Ji2015}.

We present select observational data (top panel of Fig.~12) for metal-poor DLAs ($\rm [Fe/H]<-2$) at $2.0 < z < 3.7$ (\citealp{Cooke2011}), and for higher redshift cases with $4.7 < z < 6.3$ (\citealp{Becker2012}). Intriguingly, \citet{Cooke2011} suggest that the enrichment of very metal-poor DLAs could be explained either by Pop~III CCSNe, or the yields from low-metallicity ($Z<1/3\zsun$) Pop~II SNe, based on the nitrogen abundance, which plays a critical role in distinguishing nucleosynthetic pathways. On the other hand, the DLA abundance ratios from \citet{Becker2012}, depicted as blue triangles in Fig.~12, are consistent with the integrated yields from Pop~II SNe. This implies that in the systems studied by \citet{Becker2012}, 
the Pop~III-Pop~II transition already takes place at $z>4.7$, thus in agreement with our simulation results. However, the observed DLAs may probe an environment different from our simulation, suggesting alternative metal enrichment histories.

As discussed above, over most of cosmic time, the progenitors of the simulated gas-rich dwarf galaxy, if detected, will be seen as, DLAs polluted by metals from Pop~II stars, whereas the period during which Pop~III DLAs exist in dwarfs is expected to be very brief, confined to early epochs, $z\gtrsim6$. Although note that \citet{Kulkarni2013} suggest that the imprint of Pop~III SNe on the metal ratios of DLAs will be strong enough to be detectable at $z\approx6$. In any case, our estimates of ionic column densities in Pop~III enriched minihaloes at $z\sim7$ (see Fig.~4) are similar to, or higher than, those of already observed DLAs at $z\approx6$ (\citealp{Becker2012}), reinforcing DLAs as a possible tool to study early chemical enrichment by metal-free stars. We note that next to the abundance patterns discussed in this work, abundance anomalies reported in GRB observations at $z=5.9$ (\citealp{Hartoog2015}) can also be used to diagnose the imprint of the first generation of stars.

\begin{figure}
  \includegraphics[width=100mm]{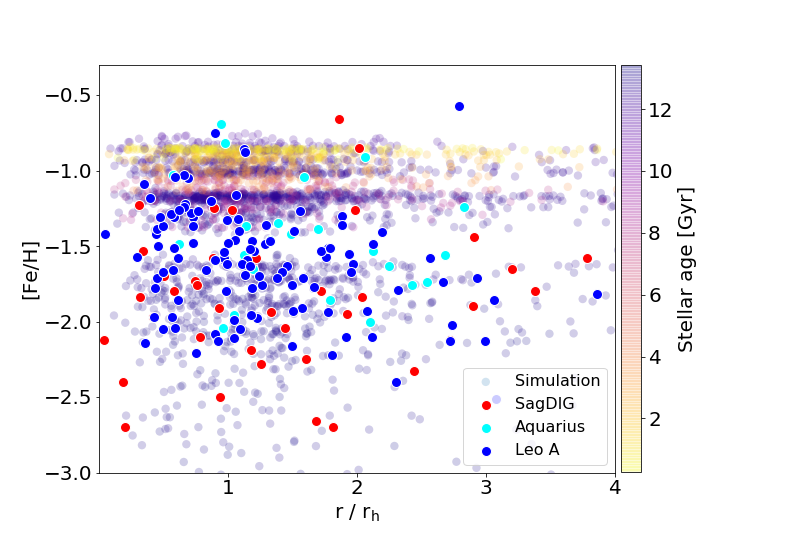}
   \caption{Stellar metallicity of member stars of the simulated dwarf (transparent background circles) as a function of radial distance normalized by half-light radius, $r_{\rm h}$, together with metallicity gradients of the observed gas-rich dwarf galaxies, SagDIG (red), Aquarius (cyan), and Leo~A (blue). The color map denotes the age of the stars in the simulated dwarf, demonstrating that overall stellar metallicity tends to rise with decreasing redshift, as one would have expected. It shows that the spread of stellar metallicity in the simulated dwarf is consistent with that of the observed dwarfs (see also \citealp{Jeon2017}).}
\end{figure}

\section{Comparison with stellar component and star formation history}
A striking difference between observed DLAs and the gas responsible for the DLA signature in our simulation, is that the simulated gas-phase metallicity reached during the assembly of the dwarf is significantly higher than what is seen in observations. To further examine whether the simulated galaxy is in reasonable agreement with observations, we compare its inferred stellar metallicities with those of member stars in the observed gas-rich, irregular dwarf galaxies SagDIG, Aquarius, and Leo~A. Their stellar masses are $1.8\times10^6\msun$, $1.5\times10^6\msun$, and $3.3\times10^6\msun$, respectively (\citealp{Kirby2017}), similar to the $M_{\ast}\approx10^6\msun$ of our simulated dwarf galaxy.

In Fig.~13, we show the distribution of stellar metallicities of the 
simulated galaxy as a function of radial distance, which is normalized by the half-light stellar radius, $r_{\rm h}=438$ pc. The colour coding denotes the stellar age, exhibiting the expected trend that stellar metallicity rises with increasing cosmic time, but old stars with stellar metallicity as high as [Fe/H]$\approx-1.2$ already exist when the age of the Universe was only 1.5~Gyr. The three dIrr systems, superimposed as filled circles, tend to host stars with a broad range of stellar metallicity, reaching from [Fe/H]$\approx-2.8$ to [Fe/H]$\approx-0.5$. Furthermore, \citet{Kirby2017} argue that the metallicity gradient of these galaxies is shallow, with a near-zero effective slope for SagDIG. We should note that other dIrrs such as Cetus or Tucana may be better analogs to the simulated galaxy, given that they appear to form more than 90$\%$ of their stars 10~Gyr ago. However, we prefer to compare with gas-rich dwarfs where high-resolution spectroscopic data is available. In addition, Cetus and Tucana seem to contain no or only little neutral gas.

\begin{figure}
  \includegraphics[width=95mm]{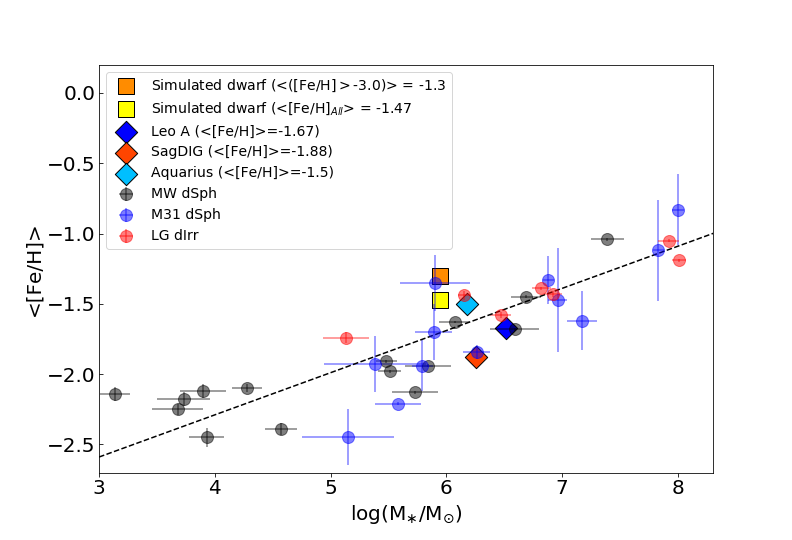}
   \caption{Mass-metallicity relation of observed MW dSph, M31 dSph, and Local Group dIrr galaxies (\citealp{Kirby2013}), together with the averaged stellar [Fe/H] in the simulated dwarf. The simulation results show reasonable agreement with the observations. We find that average stellar mass metallicity can be normal while gas phase metallicity can be high, indicating that gas phase metallicity traces current state, whereas stellar metallicity reflects gas conditions at dominant epoch of star formation.
}
\end{figure}

A noticeable difference between the simulated galaxy and the three dIrrs lies in their SFHs. We find that the simulated dwarf has formed more than $70\%$ of stars, corresponding to $M_{\ast}=6\times10^5\msun$, before $z=4$. The majority of stars thus
have ages older than 12~Gyr, while the three Local Group galaxies are late-forming dwarfs. Among them, Leo~A is the most recently forming 
dwarf with a fraction of ancient stars of less than 10$\%$ (\citealp{Cole2007}), and with an average stellar metallicity of <[Fe/H]> = -1.67 (\citealp{Kirby2017}). SagDIG is a similarly young, fast-growing dwarf with UV-measured SFR of $7.2\times10^{-4}\msun \rm yr^{-1}$ (\citealp{Karachentsev2013}), and corresponding average stellar metallicity of <[Fe/H]>=-1.88 (\citealp{Kirby2017}).

Aquarius appears to have two main populations, an ancient (> 10~Gyr) and 
a young ($\sim$ 300~Myr) one (\citealp{Ordonez2016}), with mean stellar metallicity of <[Fe/H]>\,=\,-1.5 (\citealp{Kirby2017}). The inferred SFH of Aquarius, explained by quenching of star formation in the early Universe, and the rebirth of activity at late epochs, is similar to what the simulated dwarf has experienced. However, only $\sim10\%$ of stars in Aquarius formed more than 10~Gyr ago (\citealp{Cole2014}), while in the simulated galaxy $\gtrsim70\%$ of stars had already formed prior to $z=4$. Differences in stellar metallicity could be explained by such distinct SFHs. Due to the early burst of star formation, the simulated galaxy is already metal-enriched up to <[Fe/H]>$\sim-1.2$ by $z=4$, while Aquarius likely remained metal-poor until the delayed resumption of star formation activity. Any late-forming stars in Aquarius would thus inherit its initially low gas-phase metallicity, reflected in an overall more metal-poor stellar population, compared to the simulation. Despite the detailed stellar metallicity distribution arising from different SFHs, we show that the spread of stellar metallicity in the simulated dwarf is consistent with that of the observed dwarfs (see also \citealp{Jeon2017}).

In Fig.~14, we further compare the average [Fe/H] of stars in the simulated dwarf with observed values of MW dSphs, M31 dSphs, and Local Group dIrrs, showing reasonable agreement. It is encouraging that the stellar metallicity inferred for our simulated dwarf is close to observational estimates, leading to the interpretation that its current gas-phase metallicity, which covers only a small fraction of the dwarf (see Fig.~5), is significantly enhanced above the past-averaged metallicity, as encoded in the stellar component. This explanation is in line with the fact that the simulated dwarf formed more than 70$\%$ of its stars at early times, when most gas was still metal-poor. Release of heavy elements by subsequent Type~Ia SNe continuously contributed to the total metal content of the galaxy, thus ramping up the gas-phase metallicity. Also, as the gravitational potential well becomes deeper, the galaxy more efficiently holds on to its metal-enriched gas.

\section{Summary and Conclusions}
We have explored the evolution of metal-enrichment and neutral gas content in a gas-rich dwarf galaxy, selected from the cosmological hydrodynamical simulations by \citet{Jeon2017}. We have constructed sample absorption spectra for neutral hydrogen and select metal species, associated with the high-density gas in the dwarf galaxy, comparing to observed high-density gas probed by damped Lyman-alpha (DLA) systems at high redshifts. Among the six zoom-in simulations in \citet{Jeon2017}, we have chosen the most massive, DLA harboring galaxy, with $M_{\rm halo}=4\times10^9\msun$ at $z=0$ referred to as {\sc Halo6} in \citet{Jeon2017}. As such, the pre-selected, gas-rich dwarf galaxy ($M_{\rm gas} = 10^7\msun$) maintains sufficiently high densities to produce DLA features, after surviving the disruptive effects from reionization and local stellar feedback. Below $M_{\rm halo}\lesssim3\times10^9\msun (z=0)$, simulated dwarfs tend to retain no neutral gas, which is instead removed and ionized by the external UV background and SN feedback (\citealp{Jeon2017}). This case study is thus the first detailed analysis of the chemistry and absorption line systems arising from the lowest mass halo that might plausibly contain signatures of Pop III stars in its gaseous content. 

We have found that the cross-section occupied by neutral gas, giving rise to DLAs, is very small, of order $\approx$1 kpc$^2$ at $z=0$, and further reduced to $\approx0.5$ kpc$^2$ at $z=3$. The difference is attributed to the ionization state of the gas, given the overall similar amount of gas in the dwarf at both redshifts. This indicates that detecting gas in low-mass dwarfs as DLAs is very challenging, strongly depending on the ionization state of the gas, and requiring that any observational sightline should intersect the centre of a dwarf galaxy. Moreover, if a sightline is aligned with the central region ($r$ a few $\lesssim100$ pc), the gas probed is likely to be star-forming and could thus be significantly polluted with metals, resulting in heavily saturated metal absorption lines. We also demonstrate that DLAs can only tell us about the current metallicity of their host system, and the inferred gas-phase metallicity does not need to be similar to the stellar metallicity, which probes the past-average value.

Due to the small cross-section covered by high column-density neutral gas, the velocity width associated with the DLA is expected to be small, with $\Delta v_{90}\approx 14$~km~$\rm s^{-1}$ over cosmic time from the progenitor minihalo at $z=7$ to the final dwarf at $z=0$. This causes the simulated galaxy to be offset from the empirical $\Delta v_{90}$-metallicity relation, because the central gas cloud at radii of a few $\lesssim100$~pc is highly metal-enriched. The metal enrichment history of the simulated dwarf is imprinted in the evolving chemical abundance pattern of the gas, expressed as ratios between different metal species. Specifically, gas preserving the unique signature of Pop~III stars could be discovered in the progenitor minihaloes at high redshifts, $z\gtrsim7$. Such fingerprints, however, are likely to be erased by subsequent metal enrichment from Pop~II SNe, resulting in gas clouds that mainly exhibit Pop~II abundance patterns at intermediate and recent times, in agreement with DLA observations.

This result implies that only the lowest mass dwarfs, residing in minihaloes that are both isolated and have very low merger rates, can preserve clear signatures of first generation star formation. The dwarfs that satisfy these conditions are expected to be at the extreme low-mass end of the mass function, with $M_{\rm vir}\approx10^{7-8}\msun (z=0)$, only experiencing Pop~III star formation. Furthermore, the possible observable epochs for such Pop~III DLAs might be limited to high redshifts, $z\gtrsim7$. The alternative scenario is one where the dwarf retains neutral gas after reionization through mergers or accretion, but does not trigger any subsequent star formation, such that the only contamination is from Pop III stars.  

In contrast, the simulated dwarf studied in this work has experienced early metal-enrichment, quickly forming more than $70\%$ of stars prior to $z\approx4$, corresponding to the stellar mass of $M_{\ast}=3\times10^5\msun$ by $z\approx4$. As a result, the galaxy was already composed of stars with a wide range in metallicity, from $\rm [Fe/H]\approx -5$ to as high as $\rm [Fe/H]\approx -1.2$ by $z=4$, giving rise to the average stellar metallicity of $\rm [Fe/H]\approx -1.5$ at $z=0$, in reasonable agreement with what is observed in local dwarf galaxies. Thus, the lowest mass, gas-rich dwarf galaxies observed today (Leo~A, Leo~P) are unlikely to be observable as DLAs with Pop III signatures at high redshifts. However, a larger sample size may produce a wider variation in star formation histories, and thus metal enrichment histories, than the case study presented here. As such, it remains plausible that some fraction of extremely metal-poor DLAs may be associated with low mass dwarf galaxy hosts.

DLA absorption features are a valuable tool, allowing us to investigate the conditions in the ISM and IGM at any given cosmic time. Compared to the observed metal-rich DLA systems, ascertaining the origin of metal-poor DLAs and inferring the physical nature of their associated host systems still remains challenging, primarily due to the faintness of their counterparts in emission. 
We finally note that the paucity of bright quasars at $z\gtrsim 6$ suggests an alternative way of probing cosmic metal enrichment at high-$z$, utilizing the smooth afterglow emission of gamma-ray bursts  (GRBs), which are more prevalent than quasars at $z\gtrsim 6$, as luminous background sources (e.g. \citealp{Wang2012, Ma2017}). Given that GRBs may originate inside the same galaxies that also host the dense clouds of neutral gas, effective impact parameters would be small (e.g. \citealp{Arabsalmani2015}), thus allowing the GRB afterglows to directly probe ISM conditions in high-$z$ galaxies.

\begin{acknowledgements}
We thank the anonymous referee for constructive and insightful comments that improved the clarity of our paper. We would like to thank Ryan Cooke and Yumi Choi for helpful discussions. We also thank Simeon Bird and Cameron Hummels for sharing us their packages, {\sc Fake Spectra} and {\sc Trident}. We are grateful to Volker Springel, Joop Schaye, and Claudio Dalla
Vecchia for permission to use their versions of \textsc{gadget}. G.~B. and M.~J. acknowledge support from HST Grant 15030. The simulations were performed using the El Gato cluster at the University of Arizona, which is funded by the National Science Foundation through Grant No. 1228509. We utilized \textsc{yt} for data visualization and analysis tools. M.~J. is supported by a National 
Research Foundation of Korea (NRF) grant (NRF-2018R1C1B6004304), funded by the Korean government (MSIT). V.~B. acknowledges support from NSF grant AST-1413501. The authors acknowledge the Texas Advanced 
Computing Center (TACC) at The University of Texas at Austin for providing 
HPC resources under XSEDE allocation TG-AST120024.
\end{acknowledgements}

\bibliographystyle{apj}
\bibliography{apj-jour,myrefs2}

\begin{thebibliography}{}
\expandafter\ifx\csname natexlab\endcsname\relax\def\natexlab#1{#1}\fi

\bibitem[{{Arabsalmani} {et~al.}(2015){Arabsalmani}, {M{\o}ller}, {Fynbo},
  {Christensen}, {Freudling}, {Savaglio}, \& {Zafar}}]{Arabsalmani2015}
{Arabsalmani}, M., {M{\o}ller}, P., {Fynbo}, J.~P.~U., {et~al.} 2015, MNRAS,
  446, 990

\bibitem[{{Becker} {et~al.}(2012){Becker}, {Sargent}, {Rauch}, \&
  {Carswell}}]{Becker2012}
{Becker}, G.~D., {Sargent}, W.~L.~W., {Rauch}, M., \& {Carswell}, R.~F. 2012,
  ApJ, 744, 91

\bibitem[{{Berry} {et~al.}(2016){Berry}, {Somerville}, {Gawiser}, {Maller},
  {Popping}, \& {Trager}}]{Berry2016}
{Berry}, M., {Somerville}, R.~S., {Gawiser}, E., {et~al.} 2016, MNRAS, 458, 531

\bibitem[{{Bird} {et~al.}(2015){Bird}, {Haehnelt}, {Neeleman}, {Genel},
  {Vogelsberger}, \& {Hernquist}}]{Bird2015}
{Bird}, S., {Haehnelt}, M., {Neeleman}, M., {et~al.} 2015, MNRAS, 447, 1834

\bibitem[{{Bromm}(2013)}]{Bromm2013}
{Bromm}, V. 2013, Rep. Prog. Phys., 76, 112901

\bibitem[{{Cen}(2012)}]{Cen2012}
{Cen}, R. 2012, ApJ, 748, 121

\bibitem[{{Christensen} {et~al.}(2014){Christensen}, {M{\o}ller}, {Fynbo}, \&
  {Zafar}}]{Christensen2014}
{Christensen}, L., {M{\o}ller}, P., {Fynbo}, J.~P.~U., \& {Zafar}, T. 2014,
  MNRAS, 445, 225

\bibitem[{{Cole} {et~al.}(2014){Cole}, {Weisz}, {Dolphin}, {Skillman},
  {McConnachie}, {Brooks}, \& {Leaman}}]{Cole2014}
{Cole}, A.~A., {Weisz}, D.~R., {Dolphin}, A.~E., {et~al.} 2014, ApJ, 795, 54

\bibitem[{{Cole} {et~al.}(2007){Cole}, {Skillman}, {Tolstoy}, {Gallagher},
  {Aparicio}, {Dolphin}, {Gallart}, {Hidalgo}, {Saha}, {Stetson}, \&
  {Weisz}}]{Cole2007}
{Cole}, A.~A., {Skillman}, E.~D., {Tolstoy}, E., {et~al.} 2007, ApJL, 659, L17

\bibitem[{{Cooke} {et~al.}(2011){Cooke}, {Pettini}, {Steidel}, {Rudie}, \&
  {Nissen}}]{Cooke2011}
{Cooke}, R., {Pettini}, M., {Steidel}, C.~C., {Rudie}, G.~C., \& {Nissen},
  P.~E. 2011, MNRAS, 417, 1534

\bibitem[{{Cooke} {et~al.}(2015){Cooke}, {Pettini}, \& {Jorgenson}}]{Cooke2015}
{Cooke}, R.~J., {Pettini}, M., \& {Jorgenson}, R.~A. 2015, ApJ, 800, 12

\bibitem[{{Cooke} {et~al.}(2017){Cooke}, {Pettini}, \& {Steidel}}]{Cooke2017}
{Cooke}, R.~J., {Pettini}, M., \& {Steidel}, C.~C. 2017, MNRAS,
  arXiv:1701.03103

\bibitem[{{Corlies} {et~al.}(2018){Corlies}, {Johnston}, \&
  {Wise}}]{Corlies2018}
{Corlies}, L., {Johnston}, K.~V., \& {Wise}, J.~H. 2018, MNRAS, 475, 4868

\bibitem[{{Ferland} {et~al.}(2013){Ferland}, {Porter}, {van Hoof}, {Williams},
  {Abel}, {Lykins}, {Shaw}, {Henney}, \& {Stancil}}]{Ferland2013}
{Ferland}, G.~J., {Porter}, R.~L., {van Hoof}, P.~A.~M., {et~al.} 2013, RMXAA,
  49, 137

\bibitem[{{Frebel} \& {Bromm}(2012)}]{Frebel2012}
{Frebel}, A., \& {Bromm}, V. 2012, \apj, 759, 115

\bibitem[{{Frebel} {et~al.}(2007){Frebel}, {Johnson}, \& {Bromm}}]{Frebel2007}
{Frebel}, A., {Johnson}, J.~L., \& {Bromm}, V. 2007, MNRAS, 380, L40

\bibitem[{{Fumagalli} {et~al.}(2015){Fumagalli}, {O'Meara}, {Prochaska},
  {Rafelski}, \& {Kanekar}}]{Fumagalli2015}
{Fumagalli}, M., {O'Meara}, J.~M., {Prochaska}, J.~X., {Rafelski}, M., \&
  {Kanekar}, N. 2015, MNRAS, 446, 3178

\bibitem[{{Fynbo} {et~al.}(2003){Fynbo}, {Ledoux}, {M{\o}ller}, {Thomsen}, \&
  {Burud}}]{Fynbo2003}
{Fynbo}, J.~P.~U., {Ledoux}, C., {M{\o}ller}, P., {Thomsen}, B., \& {Burud}, I.
  2003, A\&A, 407, 147

\bibitem[{{Fynbo} {et~al.}(2008){Fynbo}, {Prochaska}, {Sommer-Larsen},
  {Dessauges-Zavadsky}, \& {M{\o}ller}}]{Fynbo2008}
{Fynbo}, J.~P.~U., {Prochaska}, J.~X., {Sommer-Larsen}, J.,
  {Dessauges-Zavadsky}, M., \& {M{\o}ller}, P. 2008, ApJ, 683, 321

\bibitem[{{Fynbo} {et~al.}(2010){Fynbo}, {Laursen}, {Ledoux}, {M{\o}ller},
  {Durgapal}, {Goldoni}, {Gullberg}, {Kaper}, {Maund}, {Noterdaeme},
  {{\"O}stlin}, {Strandet}, {Toft}, {Vreeswijk}, \& {Zafar}}]{Fynbo2010}
{Fynbo}, J.~P.~U., {Laursen}, P., {Ledoux}, C., {et~al.} 2010, MNRAS, 408, 2128

\bibitem[{{Fynbo} {et~al.}(2011){Fynbo}, {Ledoux}, {Noterdaeme}, {Christensen},
  {M{\o}ller}, {Durgapal}, {Goldoni}, {Kaper}, {Krogager}, {Laursen}, {Maund},
  {Milvang-Jensen}, {Okoshi}, {Rasmussen}, {Thorsen}, {Toft}, \&
  {Zafar}}]{Fynbo2011}
{Fynbo}, J.~P.~U., {Ledoux}, C., {Noterdaeme}, P., {et~al.} 2011, MNRAS, 413,
  2481

\bibitem[{{Fynbo} {et~al.}(2013){Fynbo}, {Geier}, {Christensen}, {Gallazzi},
  {Krogager}, {Kr{\"u}hler}, {Ledoux}, {Maund}, {M{\o}ller}, {Noterdaeme},
  {Rivera-Thorsen}, \& {Vestergaard}}]{Fynbo2013}
{Fynbo}, J.~P.~U., {Geier}, S.~J., {Christensen}, L., {et~al.} 2013, MNRAS,
  436, 361

\bibitem[{{Fynbo} {et~al.}(2001){Fynbo}, {M{\o}ller}, \& {Thomsen}}]{Fynbo2001}
{Fynbo}, J.~U., {M{\o}ller}, P., \& {Thomsen}, B. 2001, A\&A, 368, 408

\bibitem[{{Fynbo} {et~al.}(1999){Fynbo}, {M{\o}ller}, \& {Warren}}]{Fynbo1999}
{Fynbo}, J.~U., {M{\o}ller}, P., \& {Warren}, S.~J. 1999, MNRAS, 305, 849

\bibitem[{{Gardner} {et~al.}(1997){Gardner}, {Katz}, {Hernquist}, \&
  {Weinberg}}]{Gardner1997}
{Gardner}, J.~P., {Katz}, N., {Hernquist}, L., \& {Weinberg}, D.~H. 1997, ApJ,
  484, 31

\bibitem[{{Gardner} {et~al.}(2001){Gardner}, {Katz}, {Hernquist}, \&
  {Weinberg}}]{Gardner2001}
---. 2001, ApJ, 559, 131

\bibitem[{{Greif} {et~al.}(2009){Greif}, {Johnson}, {Klessen}, \&
  {Bromm}}]{Greif2009}
{Greif}, T.~H., {Johnson}, J.~L., {Klessen}, R.~S., \& {Bromm}, V. 2009, MNRAS,
  399, 639

\bibitem[{{Haardt} \& {Madau}(2012)}]{Haardt2012}
{Haardt}, F., \& {Madau}, P. 2012, ApJ, 746, 125

\bibitem[{{Haehnelt} {et~al.}(1998){Haehnelt}, {Steinmetz}, \&
  {Rauch}}]{Haehnelt1998}
{Haehnelt}, M.~G., {Steinmetz}, M., \& {Rauch}, M. 1998, ApJ, 495, 647

\bibitem[{{Hartoog} {et~al.}(2015){Hartoog}, {Malesani}, {Fynbo}, {Goto},
  {Kr{\"u}hler}, {Vreeswijk}, {De Cia}, {Xu}, {M{\o}ller}, {Covino}, {D'Elia},
  {Flores}, {Goldoni}, {Hjorth}, {Jakobsson}, {Krogager}, {Kaper}, {Ledoux},
  {Levan}, {Milvang-Jensen}, {Sollerman}, {Sparre}, {Tagliaferri}, {Tanvir},
  {de Ugarte Postigo}, {Vergani}, {Wiersema}, {Datson}, {Salinas}, {Mikkelsen},
  \& {Aghanim}}]{Hartoog2015}
{Hartoog}, O.~E., {Malesani}, D., {Fynbo}, J.~P.~U., {et~al.} 2015, A\&A, 580,
  A139

\bibitem[{{Hartwig} {et~al.}(2018){Hartwig}, {Bromm}, \& {Loeb}}]{Hartwig2018}
{Hartwig}, T., {Bromm}, V., \& {Loeb}, A. 2018, MNRAS, 479, 2202

\bibitem[{{Heger} \& {Woosley}(2002)}]{Heger2002}
{Heger}, A., \& {Woosley}, S.~E. 2002, ApJ, 567, 532

\bibitem[{{Heger} \& {Woosley}(2010)}]{Heger2010}
---. 2010, ApJ, 724, 341

\bibitem[{{Hong} {et~al.}(2010){Hong}, {Katz}, {Dav{\'e}}, {Fardal}, {Kere{\v
  s}}, \& {Oppenheimer}}]{Hong2010}
{Hong}, S., {Katz}, N., {Dav{\'e}}, R., {et~al.} 2010, arXiv:1008.4242,
  arXiv:1008.4242

\bibitem[{{Hummels} {et~al.}(2017){Hummels}, {Smith}, \&
  {Silvia}}]{Hummels2017}
{Hummels}, C.~B., {Smith}, B.~D., \& {Silvia}, D.~W. 2017, ApJ, 847, 59

\bibitem[{{Jaacks} {et~al.}(2019){Jaacks}, {Finkelstein}, \&
  {Bromm}}]{Jaacks2018}
{Jaacks}, J., {Finkelstein}, S.~L., \& {Bromm}, V. 2019, arXiv:1804.07372,
  arXiv:1804.07372

\bibitem[{{Jaacks} {et~al.}(2018){Jaacks}, {Thompson}, {Finkelstein}, \&
  {Bromm}}]{Jaacks2018a}
{Jaacks}, J., {Thompson}, R., {Finkelstein}, S.~L., \& {Bromm}, V. 2018,
  \mnras, 475, 4396

\bibitem[{{Jeon} {et~al.}(2017){Jeon}, {Besla}, \& {Bromm}}]{Jeon2017}
{Jeon}, M., {Besla}, G., \& {Bromm}, V. 2017, ApJ, 848, 85

\bibitem[{{Jeon} {et~al.}(2014){Jeon}, {Pawlik}, {Bromm}, \&
  {Milosavljevic}}]{Jeon2014}
{Jeon}, M., {Pawlik}, A.~H., {Bromm}, V., \& {Milosavljevic}, M. 2014, MNRAS,
  444, 3288

\bibitem[{{Ji} {et~al.}(2015){Ji}, {Frebel}, \& {Bromm}}]{Ji2015}
{Ji}, A.~P., {Frebel}, A., \& {Bromm}, V. 2015, MNRAS, 454, 659

\bibitem[{{Jorgenson} {et~al.}(2013){Jorgenson}, {Murphy}, \&
  {Thompson}}]{Jorgenson2013}
{Jorgenson}, R.~A., {Murphy}, M.~T., \& {Thompson}, R. 2013, MNRAS, 435, 482

\bibitem[{{Jorgenson} \& {Wolfe}(2014)}]{Jorgenson2014}
{Jorgenson}, R.~A., \& {Wolfe}, A.~M. 2014, ApJ, 785, 16

\bibitem[{{Kanekar} {et~al.}(2018){Kanekar}, {Prochaska}, {Christensen},
  {Rhodin}, {Neeleman}, {Zwaan}, {M{\o}ller}, {Dessauges-Zavadsky}, {Fynbo}, \&
  {Zafar}}]{Kanekar2018}
{Kanekar}, N., {Prochaska}, J.~X., {Christensen}, L., {et~al.} 2018, ApJL, 856,
  L23

\bibitem[{{Karachentsev} \& {Kaisina}(2013)}]{Karachentsev2013}
{Karachentsev}, I.~D., \& {Kaisina}, E.~I. 2013, AJ, 146, 46

\bibitem[{{Katz} {et~al.}(1996){Katz}, {Weinberg}, {Hernquist}, \&
  {Miralda-Escude}}]{Katz1996}
{Katz}, N., {Weinberg}, D.~H., {Hernquist}, L., \& {Miralda-Escude}, J. 1996,
  ApJL, 457, L57

\bibitem[{{Kirby} {et~al.}(2013){Kirby}, {Cohen}, {Guhathakurta}, {Cheng},
  {Bullock}, \& {Gallazzi}}]{Kirby2013}
{Kirby}, E.~N., {Cohen}, J.~G., {Guhathakurta}, P., {et~al.} 2013, ApJ, 779,
  102

\bibitem[{{Kirby} {et~al.}(2017){Kirby}, {Rizzi}, {Held}, {Cohen}, {Cole},
  {Manning}, {Skillman}, \& {Weisz}}]{Kirby2017}
{Kirby}, E.~N., {Rizzi}, L., {Held}, E.~V., {et~al.} 2017, ApJ, 834, 9

\bibitem[{{Krogager} {et~al.}(2017){Krogager}, {M{\o}ller}, {Fynbo}, \&
  {Noterdaeme}}]{Krogager2017}
{Krogager}, J.-K., {M{\o}ller}, P., {Fynbo}, J.~P.~U., \& {Noterdaeme}, P.
  2017, MNRAS, 469, 2959

\bibitem[{{Krogager} {et~al.}(2013){Krogager}, {Fynbo}, {Ledoux},
  {Christensen}, {Gallazzi}, {Laursen}, {M{\o}ller}, {Noterdaeme},
  {P{\'e}roux}, {Pettini}, \& {Vestergaard}}]{Krogager2013}
{Krogager}, J.-K., {Fynbo}, J.~P.~U., {Ledoux}, C., {et~al.} 2013, MNRAS, 433,
  3091

\bibitem[{{Kulkarni} {et~al.}(2013){Kulkarni}, {Rollinde}, {Hennawi}, \&
  {Vangioni}}]{Kulkarni2013}
{Kulkarni}, G., {Rollinde}, E., {Hennawi}, J.~F., \& {Vangioni}, E. 2013, ApJ,
  772, 93

\bibitem[{{Ledoux} {et~al.}(2006){Ledoux}, {Petitjean}, {Fynbo}, {M{\o}ller},
  \& {Srianand}}]{Ledoux2006}
{Ledoux}, C., {Petitjean}, P., {Fynbo}, J.~P.~U., {M{\o}ller}, P., \&
  {Srianand}, R. 2006, A\&A, 457, 71

\bibitem[{{Ma} {et~al.}(2017){Ma}, {Maio}, {Ciardi}, \& {Salvaterra}}]{Ma2017}
{Ma}, Q., {Maio}, U., {Ciardi}, B., \& {Salvaterra}, R. 2017, MNRAS, 466, 1140

\bibitem[{{Maio} \& {Tescari}(2015)}]{Maio2015}
{Maio}, U., \& {Tescari}, E. 2015, MNRAS, 453, 3798

\bibitem[{{McConnachie}(2012)}]{McConnachie2012}
{McConnachie}, A.~W. 2012, AJ, 144, 4

\bibitem[{{M{\o}ller} {et~al.}(2004){M{\o}ller}, {Fynbo}, \&
  {Fall}}]{Moller2004}
{M{\o}ller}, P., {Fynbo}, J.~P.~U., \& {Fall}, S.~M. 2004, A\&A, 422, L33

\bibitem[{{M{\o}ller} {et~al.}(2013){M{\o}ller}, {Fynbo}, {Ledoux}, \&
  {Nilsson}}]{Moller2013}
{M{\o}ller}, P., {Fynbo}, J.~P.~U., {Ledoux}, C., \& {Nilsson}, K.~K. 2013,
  MNRAS, 430, 2680

\bibitem[{{M{\o}ller} {et~al.}(2002){M{\o}ller}, {Warren}, {Fall}, {Fynbo}, \&
  {Jakobsen}}]{Moller2002}
{M{\o}ller}, P., {Warren}, S.~J., {Fall}, S.~M., {Fynbo}, J.~U., \& {Jakobsen},
  P. 2002, ApJ, 574, 51

\bibitem[{{M{\o}ller} {et~al.}(2018){M{\o}ller}, {Christensen}, {Zwaan},
  {Kanekar}, {Prochaska}, {Rhodin}, {Dessauges-Zavadsky}, {Fynbo}, {Neeleman},
  \& {Zafar}}]{Moller2018}
{M{\o}ller}, P., {Christensen}, L., {Zwaan}, M.~A., {et~al.} 2018, MNRAS, 474,
  4039

\bibitem[{{Murphy} {et~al.}(2007){Murphy}, {Curran}, {Webb}, {M{\'e}nager}, \&
  {Zych}}]{Murphy2007}
{Murphy}, M.~T., {Curran}, S.~J., {Webb}, J.~K., {M{\'e}nager}, H., \& {Zych},
  B.~J. 2007, MNRAS, 376, 673

\bibitem[{{Nagamine} {et~al.}(2007){Nagamine}, {Wolfe}, {Hernquist}, \&
  {Springel}}]{Nagamine2007}
{Nagamine}, K., {Wolfe}, A.~M., {Hernquist}, L., \& {Springel}, V. 2007, ApJ,
  660, 945

\bibitem[{{Neeleman} {et~al.}(2013){Neeleman}, {Wolfe}, {Prochaska}, \&
  {Rafelski}}]{Neeleman2013}
{Neeleman}, M., {Wolfe}, A.~M., {Prochaska}, J.~X., \& {Rafelski}, M. 2013,
  ApJ, 769, 54

\bibitem[{{Noterdaeme} {et~al.}(2014){Noterdaeme}, {Petitjean}, {P{\^a}ris},
  {Cai}, {Finley}, {Ge}, {Pieri}, \& {York}}]{Noterdaeme2014}
{Noterdaeme}, P., {Petitjean}, P., {P{\^a}ris}, I., {et~al.} 2014, A\&A, 566,
  A24

\bibitem[{{Ordo{\~n}ez} \& {Sarajedini}(2016)}]{Ordonez2016}
{Ordo{\~n}ez}, A.~J., \& {Sarajedini}, A. 2016, MNRAS, 455, 2163

\bibitem[{{Pettini} {et~al.}(2008){Pettini}, {Zych}, {Steidel}, \&
  {Chaffee}}]{Pettini2008}
{Pettini}, M., {Zych}, B.~J., {Steidel}, C.~C., \& {Chaffee}, F.~H. 2008,
  MNRAS, 385, 2011

\bibitem[{{Pontzen} {et~al.}(2008){Pontzen}, {Governato}, {Pettini}, {Booth},
  {Stinson}, {Wadsley}, {Brooks}, {Quinn}, \& {Haehnelt}}]{Pontzen2008}
{Pontzen}, A., {Governato}, F., {Pettini}, M., {et~al.} 2008, MNRAS, 390, 1349

\bibitem[{{Portinari} {et~al.}(1998){Portinari}, {Chiosi}, \&
  {Bressan}}]{Portinari1998}
{Portinari}, L., {Chiosi}, C., \& {Bressan}, A. 1998, A\&A, 334, 505

\bibitem[{{Prochaska} {et~al.}(2008){Prochaska}, {Chen}, {Wolfe},
  {Dessauges-Zavadsky}, \& {Bloom}}]{Prochaska2008}
{Prochaska}, J.~X., {Chen}, H.-W., {Wolfe}, A.~M., {Dessauges-Zavadsky}, M., \&
  {Bloom}, J.~S. 2008, ApJ, 672, 59

\bibitem[{{Prochaska} \& {Wolfe}(1997)}]{Prochaska1997}
{Prochaska}, J.~X., \& {Wolfe}, A.~M. 1997, ApJ, 487, 73

\bibitem[{{Rafelski} {et~al.}(2012){Rafelski}, {Wolfe}, {Prochaska},
  {Neeleman}, \& {Mendez}}]{Rafelski2012}
{Rafelski}, M., {Wolfe}, A.~M., {Prochaska}, J.~X., {Neeleman}, M., \&
  {Mendez}, A.~J. 2012, ApJ, 755, 89

\bibitem[{{Rahmati} \& {Schaye}(2014)}]{Rahmati2014}
{Rahmati}, A., \& {Schaye}, J. 2014, MNRAS, 438, 529

\bibitem[{{Rauch} {et~al.}(2008){Rauch}, {Haehnelt}, {Bunker}, {Becker},
  {Marleau}, {Graham}, {Cristiani}, {Jarvis}, {Lacey}, {Morris}, {Peroux},
  {R{\"o}ttgering}, \& {Theuns}}]{Rauch2008}
{Rauch}, M., {Haehnelt}, M., {Bunker}, A., {et~al.} 2008, ApJ, 681, 856

\bibitem[{{Rhodin} {et~al.}(2018){Rhodin}, {Christensen}, {M{\o}ller}, {Zafar},
  \& {Fynbo}}]{Rhodin2018}
{Rhodin}, N.~H.~P., {Christensen}, L., {M{\o}ller}, P., {Zafar}, T., \&
  {Fynbo}, J.~P.~U. 2018, A\&A, 618, A129

\bibitem[{{Salvadori} \& {Ferrara}(2012)}]{Salvadori2012}
{Salvadori}, S., \& {Ferrara}, A. 2012, MNRAS, 421, L29

\bibitem[{{Salvadori} {et~al.}(2007){Salvadori}, {Schneider}, \&
  {Ferrara}}]{Salvadori2007}
{Salvadori}, S., {Schneider}, R., \& {Ferrara}, A. 2007, MNRAS, 381, 647

\bibitem[{{Schaye} {et~al.}(2010){Schaye}, {Dalla Vecchia}, {Booth}, {Wiersma},
  {Theuns}, {Haas}, {Bertone}, {Duffy}, {McCarthy}, \& {van de
  Voort}}]{Schaye2010}
{Schaye}, J., {Dalla Vecchia}, C., {Booth}, C.~M., {et~al.} 2010, MNRAS, 402,
  1536

\bibitem[{{Simpson} {et~al.}(2013){Simpson}, {Bryan}, {Johnston}, {Smith}, {Mac
  Low}, {Sharma}, \& {Tumlinson}}]{Simpson2013}
{Simpson}, C.~M., {Bryan}, G.~L., {Johnston}, K.~V., {et~al.} 2013, MNRAS, 432,
  1989

\bibitem[{{Springel}(2005)}]{Springel2005}
{Springel}, V. 2005, MNRAS, 364, 1105

\bibitem[{{Springel} {et~al.}(2001){Springel}, {White}, {Tormen}, \&
  {Kauffmann}}]{Springel2001}
{Springel}, V., {White}, S.~D.~M., {Tormen}, G., \& {Kauffmann}, G. 2001,
  MNRAS, 328, 726

\bibitem[{{Steidel} {et~al.}(2010){Steidel}, {Erb}, {Shapley}, {Pettini},
  {Reddy}, {Bogosavljevi{\'c}}, {Rudie}, \& {Rakic}}]{Steidel2010}
{Steidel}, C.~C., {Erb}, D.~K., {Shapley}, A.~E., {et~al.} 2010, ApJ, 717, 289

\bibitem[{{Tescari} {et~al.}(2009){Tescari}, {Viel}, {Tornatore}, \&
  {Borgani}}]{Tescari2009}
{Tescari}, E., {Viel}, M., {Tornatore}, L., \& {Borgani}, S. 2009, MNRAS, 397,
  411

\bibitem[{{Tolstoy} {et~al.}(2009){Tolstoy}, {Hill}, \& {Tosi}}]{Tolstoy2009}
{Tolstoy}, E., {Hill}, V., \& {Tosi}, M. 2009, ARA\&A, 47, 371

\bibitem[{{Umeda} \& {Nomoto}(2003)}]{Umeda2003}
{Umeda}, H., \& {Nomoto}, K. 2003, Nature, 422, 871

\bibitem[{{Verhamme} {et~al.}(2008){Verhamme}, {Schaerer}, {Atek}, \&
  {Tapken}}]{Verhamme2008}
{Verhamme}, A., {Schaerer}, D., {Atek}, H., \& {Tapken}, C. 2008, A\&A, 491, 89

\bibitem[{{Wang} {et~al.}(2012){Wang}, {Bromm}, {Greif}, {Stacy}, {Dai},
  {Loeb}, \& {Cheng}}]{Wang2012}
{Wang}, F.~Y., {Bromm}, V., {Greif}, T.~H., {et~al.} 2012, ApJ, 760, 27

\bibitem[{{Webster} {et~al.}(2015){Webster}, {Bland-Hawthorn}, \&
  {Sutherland}}]{Webster2015b}
{Webster}, D., {Bland-Hawthorn}, J., \& {Sutherland}, R.~S. 2015, ApJ, 804, 110

\bibitem[{{Whalen} {et~al.}(2013){Whalen}, {Even}, {Frey}, {Smidt}, {Johnson},
  {Lovekin}, {Fryer}, {Stiavelli}, {Holz}, {Heger}, {Woosley}, \&
  {Hungerford}}]{Whalen2013}
{Whalen}, D.~J., {Even}, W., {Frey}, L.~H., {et~al.} 2013, ApJ, 777, 110

\bibitem[{{Wise} {et~al.}(2012){Wise}, {Turk}, {Norman}, \& {Abel}}]{Wise2012}
{Wise}, J.~H., {Turk}, M.~J., {Norman}, M.~L., \& {Abel}, T. 2012, ApJ, 745, 50

\bibitem[{{Wolfe} {et~al.}(2005){Wolfe}, {Gawiser}, \& {Prochaska}}]{Wolfe2005}
{Wolfe}, A.~M., {Gawiser}, E., \& {Prochaska}, J.~X. 2005, ARA\&A, 43, 861

\bibitem[{{Wolfe} {et~al.}(1986){Wolfe}, {Turnshek}, {Smith}, \&
  {Cohen}}]{Wolfe1986}
{Wolfe}, A.~M., {Turnshek}, D.~A., {Smith}, H.~E., \& {Cohen}, R.~D. 1986,
  ApJS, 61, 249

\bibitem[{{Wolfire} {et~al.}(2003){Wolfire}, {McKee}, {Hollenbach}, \&
  {Tielens}}]{Wolfire2003}
{Wolfire}, M.~G., {McKee}, C.~F., {Hollenbach}, D., \& {Tielens}, A.~G.~G.~M.
  2003, ApJ, 587, 278

\bibitem[{{Yuan} \& {Cen}(2016)}]{Yuan2016}
{Yuan}, S., \& {Cen}, R. 2016, MNRAS, 457, 487

\bibitem[{{Zafar} {et~al.}(2017){Zafar}, {M{\o}ller}, {P{\'e}roux}, {Quiret},
  {Fynbo}, {Ledoux}, \& {Deharveng}}]{Zafar2017}
{Zafar}, T., {M{\o}ller}, P., {P{\'e}roux}, C., {et~al.} 2017, MNRAS, 465, 1613

\end{thebibliography}

\end{document}